\documentclass[aps,prc,superscriptaddress,showpacs,nofootinbib,floatfix,twocolumn]{revtex4}

\usepackage{graphicx}	

\begin{document}


\title{Charged hadron multiplicity fluctuations in Au+Au and Cu+Cu collisions \\
from $\sqrt{s_{\rm NN}}$ = 22.5 to 200 GeV}

\newcommand{\abilene}{Abilene Christian University, Abilene, TX 79699, USA}
\newcommand{\acadsin}{Institute of Physics, Academia Sinica, Taipei 11529, Taiwan}
\newcommand{\banaras}{Department of Physics, Banaras Hindu University, Varanasi 221005, India}
\newcommand{\barc}{Bhabha Atomic Research Centre, Bombay 400 085, India}
\newcommand{\bnl}{Brookhaven National Laboratory, Upton, NY 11973-5000, USA}
\newcommand{\caucr}{University of California - Riverside, Riverside, CA 92521, USA}
\newcommand{\charlesczech}{Charles University, Ovocn\'{y} trh 5, Praha 1, 116 36, Prague, Czech Republic}
\newcommand{\ciae}{China Institute of Atomic Energy (CIAE), Beijing, People's Republic of China}
\newcommand{\cns}{Center for Nuclear Study, Graduate School of Science, University of Tokyo, 7-3-1 Hongo, Bunkyo, Tokyo 113-0033, Japan}
\newcommand{\colorado}{University of Colorado, Boulder, CO 80309, USA}
\newcommand{\columbia}{Columbia University, New York, NY 10027 and Nevis Laboratories, Irvington, NY 10533, USA}
\newcommand{\czechtech}{Czech Technical University, Zikova 4, 166 36 Prague 6, Czech Republic}
\newcommand{\dapnia}{Dapnia, CEA Saclay, F-91191, Gif-sur-Yvette, France}
\newcommand{\debrecen}{Debrecen University, H-4010 Debrecen, Egyetem t{\'e}r 1, Hungary}
\newcommand{\elte}{ELTE, E{\"o}tv{\"o}s Lor{\'a}nd University, H - 1117 Budapest, P{\'a}zm{\'a}ny P. s. 1/A, Hungary}
\newcommand{\fit}{Florida Institute of Technology, Melbourne, FL 32901, USA}
\newcommand{\fsu}{Florida State University, Tallahassee, FL 32306, USA}
\newcommand{\gsu}{Georgia State University, Atlanta, GA 30303, USA}
\newcommand{\hiroshima}{Hiroshima University, Kagamiyama, Higashi-Hiroshima 739-8526, Japan}
\newcommand{\ihepprot}{IHEP Protvino, State Research Center of Russian Federation, Institute for High Energy Physics, Protvino, 142281, Russia}
\newcommand{\illuiuc}{University of Illinois at Urbana-Champaign, Urbana, IL 61801, USA}
\newcommand{\instpasczech}{Institute of Physics, Academy of Sciences of the Czech Republic, Na Slovance 2, 182 21 Prague 8, Czech Republic}
\newcommand{\isu}{Iowa State University, Ames, IA 50011, USA}
\newcommand{\jinrdubna}{Joint Institute for Nuclear Research, 141980 Dubna, Moscow Region, Russia}
\newcommand{\kaeri}{KAERI, Cyclotron Application Laboratory, Seoul, Korea}
\newcommand{\kek}{KEK, High Energy Accelerator Research Organization, Tsukuba, Ibaraki 305-0801, Japan}
\newcommand{\kfki}{KFKI Research Institute for Particle and Nuclear Physics of the Hungarian Academy of Sciences (MTA KFKI RMKI), H-1525 Budapest 114, POBox 49, Budapest, Hungary}
\newcommand{\korea}{Korea University, Seoul, 136-701, Korea}
\newcommand{\kurchatov}{Russian Research Center ``Kurchatov Institute", Moscow, Russia}
\newcommand{\kyoto}{Kyoto University, Kyoto 606-8502, Japan}
\newcommand{\labllr}{Laboratoire Leprince-Ringuet, Ecole Polytechnique, CNRS-IN2P3, Route de Saclay, F-91128, Palaiseau, France}
\newcommand{\lawllnl}{Lawrence Livermore National Laboratory, Livermore, CA 94550, USA}
\newcommand{\losalamos}{Los Alamos National Laboratory, Los Alamos, NM 87545, USA}
\newcommand{\lpc}{LPC, Universit{\'e} Blaise Pascal, CNRS-IN2P3, Clermont-Fd, 63177 Aubiere Cedex, France}
\newcommand{\lund}{Department of Physics, Lund University, Box 118, SE-221 00 Lund, Sweden}
\newcommand{\muenster}{Institut f\"ur Kernphysik, University of Muenster, D-48149 Muenster, Germany}
\newcommand{\myongji}{Myongji University, Yongin, Kyonggido 449-728, Korea}
\newcommand{\nagasaki}{Nagasaki Institute of Applied Science, Nagasaki-shi, Nagasaki 851-0193, Japan}
\newcommand{\newmex}{University of New Mexico, Albuquerque, NM 87131, USA }
\newcommand{\nmsu}{New Mexico State University, Las Cruces, NM 88003, USA}
\newcommand{\ornl}{Oak Ridge National Laboratory, Oak Ridge, TN 37831, USA}
\newcommand{\orsay}{IPN-Orsay, Universite Paris Sud, CNRS-IN2P3, BP1, F-91406, Orsay, France}
\newcommand{\peking}{Peking University, Beijing, People's Republic of China}
\newcommand{\pnpi}{PNPI, Petersburg Nuclear Physics Institute, Gatchina, Leningrad region, 188300, Russia}
\newcommand{\riken}{RIKEN, The Institute of Physical and Chemical Research, Wako, Saitama 351-0198, Japan}
\newcommand{\rikjrbrc}{RIKEN BNL Research Center, Brookhaven National Laboratory, Upton, NY 11973-5000, USA}
\newcommand{\rikkyo}{Physics Department, Rikkyo University, 3-34-1 Nishi-Ikebukuro, Toshima, Tokyo 171-8501, Japan}
\newcommand{\saispbstu}{Saint Petersburg State Polytechnic University, St. Petersburg, Russia}
\newcommand{\saopaulo}{Universidade de S{\~a}o Paulo, Instituto de F\'{\i}sica, Caixa Postal 66318, S{\~a}o Paulo CEP05315-970, Brazil}
\newcommand{\seoulnat}{System Electronics Laboratory, Seoul National University, Seoul, Korea}
\newcommand{\stonybrkc}{Chemistry Department, Stony Brook University, Stony Brook, SUNY, NY 11794-3400, USA}
\newcommand{\stonycrkp}{Department of Physics and Astronomy, Stony Brook University, SUNY, Stony Brook, NY 11794, USA}
\newcommand{\subatech}{SUBATECH (Ecole des Mines de Nantes, CNRS-IN2P3, Universit{\'e} de Nantes) BP 20722 - 44307, Nantes, France}
\newcommand{\tenn}{University of Tennessee, Knoxville, TN 37996, USA}
\newcommand{\titech}{Department of Physics, Tokyo Institute of Technology, Oh-okayama, Meguro, Tokyo 152-8551, Japan}
\newcommand{\tsukuba}{Institute of Physics, University of Tsukuba, Tsukuba, Ibaraki 305, Japan}
\newcommand{\vandy}{Vanderbilt University, Nashville, TN 37235, USA}
\newcommand{\waseda}{Waseda University, Advanced Research Institute for Science and Engineering, 17 Kikui-cho, Shinjuku-ku, Tokyo 162-0044, Japan}
\newcommand{\weizmann}{Weizmann Institute, Rehovot 76100, Israel}
\newcommand{\yonsei}{Yonsei University, IPAP, Seoul 120-749, Korea}
\affiliation{\abilene}
\affiliation{\acadsin}
\affiliation{\banaras}
\affiliation{\barc}
\affiliation{\bnl}
\affiliation{\caucr}
\affiliation{\charlesczech}
\affiliation{\ciae}
\affiliation{\cns}
\affiliation{\colorado}
\affiliation{\columbia}
\affiliation{\czechtech}
\affiliation{\dapnia}
\affiliation{\debrecen}
\affiliation{\elte}
\affiliation{\fit}
\affiliation{\fsu}
\affiliation{\gsu}
\affiliation{\hiroshima}
\affiliation{\ihepprot}
\affiliation{\illuiuc}
\affiliation{\instpasczech}
\affiliation{\isu}
\affiliation{\jinrdubna}
\affiliation{\kaeri}
\affiliation{\kek}
\affiliation{\kfki}
\affiliation{\korea}
\affiliation{\kurchatov}
\affiliation{\kyoto}
\affiliation{\labllr}
\affiliation{\lawllnl}
\affiliation{\losalamos}
\affiliation{\lpc}
\affiliation{\lund}
\affiliation{\muenster}
\affiliation{\myongji}
\affiliation{\nagasaki}
\affiliation{\newmex}
\affiliation{\nmsu}
\affiliation{\ornl}
\affiliation{\orsay}
\affiliation{\peking}
\affiliation{\pnpi}
\affiliation{\riken}
\affiliation{\rikjrbrc}
\affiliation{\rikkyo}
\affiliation{\saispbstu}
\affiliation{\saopaulo}
\affiliation{\seoulnat}
\affiliation{\stonybrkc}
\affiliation{\stonycrkp}
\affiliation{\subatech}
\affiliation{\tenn}
\affiliation{\titech}
\affiliation{\tsukuba}
\affiliation{\vandy}
\affiliation{\waseda}
\affiliation{\weizmann}
\affiliation{\yonsei}
\author{A.~Adare}	\affiliation{\colorado}
\author{S.S.~Adler}	\affiliation{\bnl}
\author{S.~Afanasiev}	\affiliation{\jinrdubna}
\author{C.~Aidala}	\affiliation{\columbia}
\author{N.N.~Ajitanand}	\affiliation{\stonybrkc}
\author{Y.~Akiba}	\affiliation{\kek}  \affiliation{\riken}  \affiliation{\rikjrbrc}
\author{H.~Al-Bataineh}	\affiliation{\nmsu}
\author{J.~Alexander}	\affiliation{\stonybrkc}
\author{A.~Al-Jamel}	\affiliation{\nmsu}
\author{K.~Aoki}	\affiliation{\kyoto} \affiliation{\riken}
% ...
\collaboration{PHENIX Collaboration} \noaffiliation

\date{\today}

\begin{abstract}

A comprehensive survey of event-by-event fluctuations of charged hadron 
multiplicity in relativistic heavy ions is presented. The survey covers 
Au+Au collisions at $\sqrt{s_{\rm NN}}$ = 62.4 and 200 GeV, and Cu+Cu 
collisions at $\sqrt{s_{\rm NN}}$ = 22.5, 62.4, and 200 GeV. 
Fluctuations are measured as a function of collision centrality, 
transverse momentum range, and charge sign. After correcting for 
non-dynamical fluctuations due to fluctuations in the collision geometry 
within a centrality bin, the remaining dynamical fluctuations expressed 
as the variance normalized by the mean tend to decrease with increasing 
centrality. The dynamical fluctuations are consistent with or below the 
expectation from a superposition of participant nucleon-nucleon 
collisions based upon p+p data, indicating that this dataset does not 
exhibit evidence of critical behavior in terms of the compressibility of 
the system. A comparison of the data with a model where hadrons are 
independently emitted from a number of hadron clusters suggests that the 
mean number of hadrons per cluster is small in heavy ion collisions.

\end{abstract}

	
\maketitle

\section{Introduction}

Recent work with lattice gauge theory simulations has attempted to map 
out the phase diagram of Quantum Chromodynamics (QCD) in temperature and 
baryo-chemical potential ($\mu_{\rm B}$) using finite values of the up 
and down quark masses.  The results of these studies indicate that the 
QCD phase diagram may contain a first-order transition line between the 
hadron gas phase and the strongly-coupled Quark-Gluon Plasma (sQGP) 
phase that terminates at a critical point \cite{QCDstephanov}. This 
property is analogous to that observed in the phase diagram for many 
common liquids and other substances, including water. However, different 
model predictions and lattice calculations yield widely varying 
estimates of the location of the critical point on the QCD phase diagram 
\cite{QCDstephPre}. Direct experimental observation of critical 
phenomena in heavy ion collisions would confirm the existence of the 
critical point, narrow down its location on the QCD phase diagram, and 
provide an important constraint for the QCD models.

The estimated value of energy densities achieved in heavy ion collisions 
at the Brookhaven National Laboratory's Relativistic Heavy Ion Collider 
(RHIC) exceeds the threshold for a phase transition from normal hadronic 
matter to partonic matter.  Recent experimental evidence indicates that 
properties of the matter being produced include strong collective flow 
and large opacity to scattered quarks and gluons - the matter appears to 
behave much like a perfect fluid \cite{whitePaper}. While measurements 
suggest the produced matter has properties that differ from normal 
nuclear matter, unambiguous evidence of the nature and location of any 
phase transition from normal nuclear matter has been elusive thus far. 
Described here is a search for direct evidence of a phase transition by 
measuring the fluctuations of the event-by-event multiplicities of 
produced charge particles in a variety of collision systems.

In order to illustrate how the measurement of charged particle 
multiplicity fluctuations can be sensitive to the presence of a phase 
transition, the isothermal compressibility of the system can be 
considered \cite{Stan98}. The isothermal compressibility is defined as 
follows:
\begin{equation} \label{eq:kT}
  k_T = -1/V\left(\delta{V}/\delta{P}\right)_{\rm T},
\end{equation}

where V is the volume, T is the temperature, and P is the pressure of 
the system. In order to relate the compressibility to measurements of 
multiplicity fluctuations, we assume that relativistic nucleus-nucleus 
collisions can be described as a thermal system in the Grand Canonical 
Ensemble (GCE) \cite{Begun05}.  The GCE can be applied to the case of 
measurements near mid-rapidity since energy and conserved quantum 
numbers in this region can be exchanged with the rest of the system, 
that serves as a heat bath \cite{jeonReview}. Detailed studies of 
multiplicity fluctuations in the Canonical and Microcanonical Ensembles 
with the application of conservation laws can be found elsewhere 
\cite{Begun04,Becattini05}. In the GCE, the isothermal compressibility 
is directly related to the variance of the particle multiplicity as 
follows:
\begin{equation} \label{eq:varkt}
  \langle (N-\langle N\rangle)^2 \rangle = var(N) 
   = \frac{k_B T \langle N\rangle ^2}{V} k_T,
\end{equation}
where N is the particle multiplicity, $\langle N \rangle = \mu_N$ is the 
mean multiplicity, and $k_{\rm B}$ is Boltzmann's constant 
\cite{Stanley}.  Here, multiplicity fluctuation measurements are 
presented in terms of the scaled variance, $\omega_{\rm N}$:
\begin{equation} \label{eq:omega}
  \omega_{\rm N} = \frac{var(N)}{\mu_{\rm N}}
    = k_B T \frac{\mu_{\rm N}}{V} k_T
\end{equation}
In a continuous, or second-order, phase transition, the compressibility 
diverges to an infinite value at the critical point. Near the critical 
point, this divergence is described by a power law in the variable 
$\epsilon = (T-T_{\rm C})/T_{\rm C}$, where $T_{\rm C}$ is the critical 
temperature. Hence, the relationship between multiplicity fluctuations 
and the compressibility can be exploited to search for a clear signature 
of critical behavior by looking for the expected power law scaling of 
the compressibility:
\begin{equation} \label{eq:critExpo}
   k_{\rm T} \propto (\frac{T-T_C}{T_C})^{-\gamma} \propto \epsilon^{-\gamma},
\end{equation}
where $\gamma$ is the critical exponent for isothermal compressibility 
\cite{Stanley}. If the QCD phase diagram contains a critical point, 
systems with a low value of baryo-chemical potential ($\mu_{\rm B}$) 
could pass through the cross-over region and undergo a continuous phase 
transition \cite{QCDstephPre}. Recent estimates 
\cite{QCDsuscept1,QCDsuscept2} of the behavior of the quark number 
susceptibility, $\chi_{\rm q}$, which is proportional to the value of 
the isothermal compressibility of the system, predict that its value 
will increase by at least an order of magnitude close to the QCD 
critical point. Given that the scaled variance is proportional to 
$k_{\rm T}$, measurements of charged particle multiplicity are expected 
to be a sensitive probe for critical behavior. In addition, within a 
scenario where droplets of Quark-Gluon Plasma are formed during a 
first-order phase transition, the scaled variance of the multiplicity 
could increase by a factor of 10-100 \cite{droplets}.

Experimentally, a search for critical behavior is facilitated by the 
rich and varied dataset provided by RHIC. It is expected that the 
trajectory of the colliding system in the QCD phase diagram can be 
modified by varying the colliding energy \cite{QCDstephPre}. If the 
system approaches close enough to the critical line for a long enough 
time period, then critical phenomena could be readily apparent through 
the measurement of multiplicity fluctuations \cite{stephRajShur}. It may 
also be possible to determine the critical exponents of the system. 
Nature tends to group materials into universality classes whereby all 
materials in the same universality class share identical values for 
their set of critical exponents.  Although beyond the scope of this 
analysis, observation of critical behavior in heavy ion collisions and 
the subsequent measurement of the critical exponents could determine the 
universality class in which QCD is grouped, providing essential 
constraints for the models.

Charged particle multiplicity fluctuations have been measured in 
elementary collisions over a large range of collisions energies 
\cite{kafka75,thome77,ua5-85,ua5Clan,emcClan,na22Clan,ua5200GeV}. 
Initial measurements of multiplicity fluctuations in minimum-bias O+Cu 
collisions at $\sqrt{s_{\rm NN}}$=4.86 GeV were made by BNL Experiment 
E802 \cite{e802MF}, minimum-bias O+Au collisions at $\sqrt{s_{\rm 
NN}}$=17.3 GeV by CERN Experiment WA80 \cite{wa80MF} and minimum-bias 
S+S, O+Au, and S+Au collisions at $\sqrt{s_{\rm NN}}$=17.3 GeV by CERN 
Experiment NA35 \cite{na35Clan}. Recently, larger datasets have enabled 
the measurement of the centrality-dependence of multiplicity 
fluctuations in Pb+Pb collisions at $\sqrt{s_{\rm NN}}$=17.3 GeV by CERN 
Experiment WA98 \cite{wa98MF} and in Pb+Pb, C+C, and Si+Si collisions at 
$\sqrt{s_{\rm NN}}$=17.3 GeV by CERN Experiment NA49 \cite{na49MF}. The 
PHENIX Experiment at RHIC has performed an analysis of density 
correlations in longitudinal space with a differential analysis of 
charged particle multiplicity fluctuations in 200 GeV Au+Au collisions 
over the entire transverse momentum range \cite{ppg061}. Thus far, the 
fluctuation measurements in heavy ion collisions do not indicate 
significant signs of a phase transition. However, the full range of 
collision energies and species accessible by RHIC are yet to be 
explored.

Presented here is a comprehensive survey of multiplicity fluctuations of 
charged hadrons measured by the PHENIX Experiment at RHIC. The survey 
will cover the following collision systems: $\sqrt{s_{\rm NN}}$=200 GeV 
Au+Au, 62.4 GeV Au+Au, 200 GeV Cu+Cu, 62.4 GeV Cu+Cu, and 22.5 GeV Cu+Cu 
with comparisons to $\sqrt{s}$=200 GeV p+p collisions, which serve as a 
baseline measurement.  The Au+Au data were taken during RHIC Run-4 
(2004), the Cu+Cu data were taken during RHIC Run-5 (2005), and the p+p 
data were taken during RHIC Run-3 (2003). Multiplicity fluctuations for 
each collision system with the exception of p+p will also be presented 
as a function of centrality to help select the system volume. 
Multiplicity fluctuations will also be presented as a function of 
transverse momentum range, and charge sign.

This paper is organized as follows: Sec. II will discuss the 
experimental apparatus and details; Sec. III will discuss the methods 
applied for the measurement of multiplicity fluctuations and the removal 
of non-dynamical fluctuations due to fluctuations of the collision 
geometry within a centrality bin; Sec. IV will present the results and 
compare them to other models. Sec. V will present a discussion and 
summary of the results.

\section{Experimental Setup}

The PHENIX detector consists of two central spectrometer arms designed 
for charged particle tracking, designated east and west, and two muon 
spectrometers designed for muon tracking and identification, designated 
north and south.  The muon spectrometers are not used in this analysis. 
A comprehensive description of the PHENIX detector is documented 
elsewhere \cite{nimPHENIX}. The analysis described here utilizes the 
central spectrometer arms, which consist of a set of tracking detectors 
\cite{nimTracking}, particle identification detectors \cite{nimPID}, and 
an electromagnetic calorimeter \cite{nimEMC}. The central spectrometer 
arms cover a rapidity range of $|\eta|<0.35$ and each arm subtends 90 
degrees in azimuth. A detailed description of the algorithms and 
performance of the central arm track reconstruction and momentum 
reconstruction can be found in \cite{nimReco}.

There are two detectors that are used for triggering, centrality 
determination, and event vertex determination. The Beam-Beam Counters 
(BBCs) consist of 64 individual quartz Cherenkov counters that cover the 
full azimuthal angle in the pseudorapidity range $3.0<|\eta|<3.9$. The 
Zero Degree Calorimeters (ZDCs) cover the pseudorapidity range 
$|\eta|>6$ and measure the energy of spectator neutrons with an energy 
resolution of approximately 20\%. More details about these detectors can 
be found in \cite{nimInner}. The collision vertex position is determined 
using timing information from the BBCs with an r.m.s. resolution for 
central Au+Au events of 6 mm along the beam axis. The collision vertex 
is required to be reconstructed within $\pm$30 cm from the center of the 
spectrometer. The BBCs also provide a minimum biased (MB) event trigger.

Due to the large dynamic range in $\sqrt{s_{\rm NN}}$ covered by this 
analysis, it is necessary to implement algorithms that are dependent on 
the collision energy for the determination of the centrality of each 
event. In Au+Au collisions at $\sqrt{s_{\rm NN}}$=200 GeV, the 
centrality of the collision is determined by using correlations of the 
total energy deposited in the ZDCs with the total charge deposited in 
the BBCs as described in \cite{phxMult}. However, in 200 GeV Cu+Cu, 62.4 
GeV Cu+Cu, and 62.4 GeV Au+Au collisions, the resolving power of the 
ZDCs is insufficient to significantly contribute to the centrality 
definition. Therefore, only the total charge deposited in the BBCs is 
used to determine centrality in these collision systems, as described in 
\cite{phxMult}. Using the 200 GeV Au+Au data, it has been verified that 
application of the BBC-ZDC correlation for the centrality definition as 
opposed to the BBC-only definition shows no significant differences in 
the values of the charged hadron fluctuation quantities presented here 
as a function of centrality.

The location of the BBCs are fixed for every collision energy. At the 
lowest collision energy ($\sqrt{s_{\rm NN}}$=22.5 GeV), it becomes 
kinematically possible for spectator nucleons to fall within the 
acceptance of the BBC. This results in a BBC response in its total 
charge sum that is no longer linear with the number of participating 
nucleons ($N_{\rm part}$). In this case, it becomes necessary to define 
the centrality using the total charged particle multiplicity in Pad 
Chamber 1 (PC1) \cite{nimTracking}. PC1 is chosen due to its fine 
segmentation, high tracking efficiency, and relative proximity to the 
event vertex. Details on this procedure are also described in 
\cite{phxMult}. For all collision species and energies, the distribution 
of the number of participants was determined using a Monte Carlo 
simulation based upon the Glauber model \cite{glauber,phxMult}.

The number of minimum bias events analyzed for each dataset are 25.6 
million events for 200 GeV Au+Au, 24.9 million events for 62.4 GeV 
Au+Au, 15.0 million events for 200 GeV Cu+Cu, 12.2 million events for 
62.4 GeV Cu+Cu, 5.5 million events for 22.5 GeV Cu+Cu, and 2.7 million 
events for 200 GeV p+p.  Only a fraction of the complete 200 GeV Au+Au, 
Cu+Cu, and p+p datasets are analyzed, but this fraction is more than 
sufficient for this analysis.

The charged particle multiplicity is determined on an event-by-event 
basis by counting the number of unambiguous reconstructed tracks in the 
Drift Chamber originating from the collision vertex that have 
corresponding hits in Pad Chamber 1 and Pad Chamber 3. Track selection 
includes cuts on reconstructed tracks in the Drift Chamber to reduce 
double-counted ghost tracks to a negligible level. In order to minimize 
background originating from the magnets, reconstructed tracks are 
required to lie within $\pm$75 cm from the center of the Drift Chamber 
along the beam axis. This requirement reduces the pseudorapidity range 
of reconstructed tracks to $|\eta|<$0.26. The Ring Imaging Cherenkov 
detector (RICH) is utilized to reduce background from electrons 
resulting from photon conversions.

Although the central arm spectrometer covers a total azimuthal range of 
$\pi$ radians, detector and tracking inefficiencies reduce the effective 
average azimuthal active area to 2.1 radians for the 200 Gev Au+Au and 
200 GeV p+p datasets, and 2.0 radians for the other datasets.  
Fluctuation quantities are quoted for these acceptances separately for 
each dataset. The differences in acceptance between datasets, which are 
due to variations in the detector over the three year period in which 
the data was collected, result in less than a 1\% variation in the 
fluctuation quantities quoted here.

\section{Data Analysis}

Multiplicity fluctuations of charged particles, designated 
$\omega_{\rm ch}$, can be generally defined \cite{heiselReview} as 
follows:
\begin{equation} \label{eq:omega2}
   \omega_{\rm ch} = \frac{(\langle N_{\rm ch}^2 \rangle-\langle N_{\rm ch} \rangle^2)}{\langle N_{\rm ch} \rangle} = \frac{\sigma^{2}_{\rm ch}}{\mu_{\rm ch}},
\end{equation}
where $N_{\rm ch}$ is the charged particle multiplicity. Simply 
stated, the fluctuations can be quoted as the variance of the 
multiplicity ($\sigma_{\rm ch}^2$) normalized by the mean ($\mu_{\rm 
ch} = \langle N_{\rm ch} \rangle$). This is also referred to as the 
scaled variance \cite{na49MF}. If the multiplicity distribution is 
Poissonian, the scaled variance is 1.0.

It has been well established that charged particle multiplicity 
distributions in elementary nucleon-nucleon collisions can be 
described by the Negative Binomial Distribution (NBD) 
\cite{ua5Clan,emcClan,na22Clan}. The NBD also well describes 
multiplicity distributions in heavy ion collisions 
\cite{e802MF,na35Clan}. The Negative Binomial Distribution of an 
integer $n$ is defined as follows:
\begin{equation} \label{eq:nbd}
P(n) = \frac{\Gamma(n+k_{\rm NBD})}{\Gamma(n+1)\Gamma(k_{\rm NBD})} \frac{(\mu_{\rm ch}/k_{\rm NBD})^n}{(1+\mu_{\rm ch}/k_{\rm NBD})^{n+k_{\rm NBD}}},
\end{equation}
where $P(n)$ is normalized to 1.0 over the range 
$0 \le n \le \infty$, $\mu_{\rm ch} 
= \langle N_{\rm ch} \rangle = \langle n \rangle$, 
and $k_{\rm NBD}$ is an additional parameter. The NBD 
reduces to a Poisson distribution in the limit $k_{\rm NBD} 
\rightarrow \infty$. The NBD variance and mean is related 
to $k_{\rm NBD}$ as follows:
\begin{equation} \label{eq:nbdk}
   \frac{\sigma^{2}_{\rm ch}}{\mu^{2}_{\rm ch}} = \frac{\omega_{\rm ch}}{\mu_{\rm ch}} 
   = \frac{1}{\mu_{\rm ch}} + \frac{1}{k_{\rm NBD}}.
\end{equation}
Hence, the scaled variance is given by
\begin{equation} \label{eq:omegak}
   \omega_{\rm ch} = 1 + \frac{\mu_{\rm ch}}{k_{\rm NBD}}.
\end{equation}

A useful property of the Negative Binomial Distribution concerns its 
behavior when a population that follows the NBD is subdivided 
randomly by repeated independent trials with a constant probability 
onto smaller subsets. This results in a binomial decomposition of 
the original population into subsets that also follow the NBD with 
the same value of $k_{\rm NBD}$ \cite{e802MF}. This property can be 
applied to estimate the behavior of multiplicity fluctuations as a 
function of acceptance, assuming that there are no significant 
correlations present over the acceptance range being examined. 
Starting with an original NBD sample with mean $\mu_{\rm ch}$ and 
scaled variance $\omega_{\rm ch}$, a sample in a fractional 
acceptance with mean $\mu_{\rm acc}$ is also described by an NBD 
distribution. An acceptance fraction can be defined as 
$f_{\rm acc} = \mu_{\rm acc}/\mu_{\rm ch}$.  The scaled variance 
of the subsample from Equation (\ref{eq:omegak}) is thus
\begin{equation} \label{eq:omegaacc}
  \omega_{\rm acc} = 1 + (\mu_{\rm acc}/k_{\rm NBD}) 
  = 1 + (f_{\rm acc}\mu_{\rm ch}/k_{\rm NBD}). 
\end{equation}
Since $k_{\rm NBD}$ is identical for the two samples, $\mu_{\rm 
ch}/k_{\rm NBD} = \omega_{\rm ch} - 1$ can be substituted, yielding 
the following relation between the scaled variances of the original 
and fractional acceptance samples:
\begin{equation} \label{eq:acc}
   \omega_{\rm acc} = 1 + f_{\rm acc}(\omega_{\rm ch} - 1)
\end{equation}
Thus, the measured scaled variance will decrease as the acceptance 
is decreased while $k_{\rm NBD}$ remains constant, if there are no 
additional correlations present over the given acceptance range.

\begin{figure}[th]
\vspace{-0.2cm}
\includegraphics[width=1.0\linewidth]{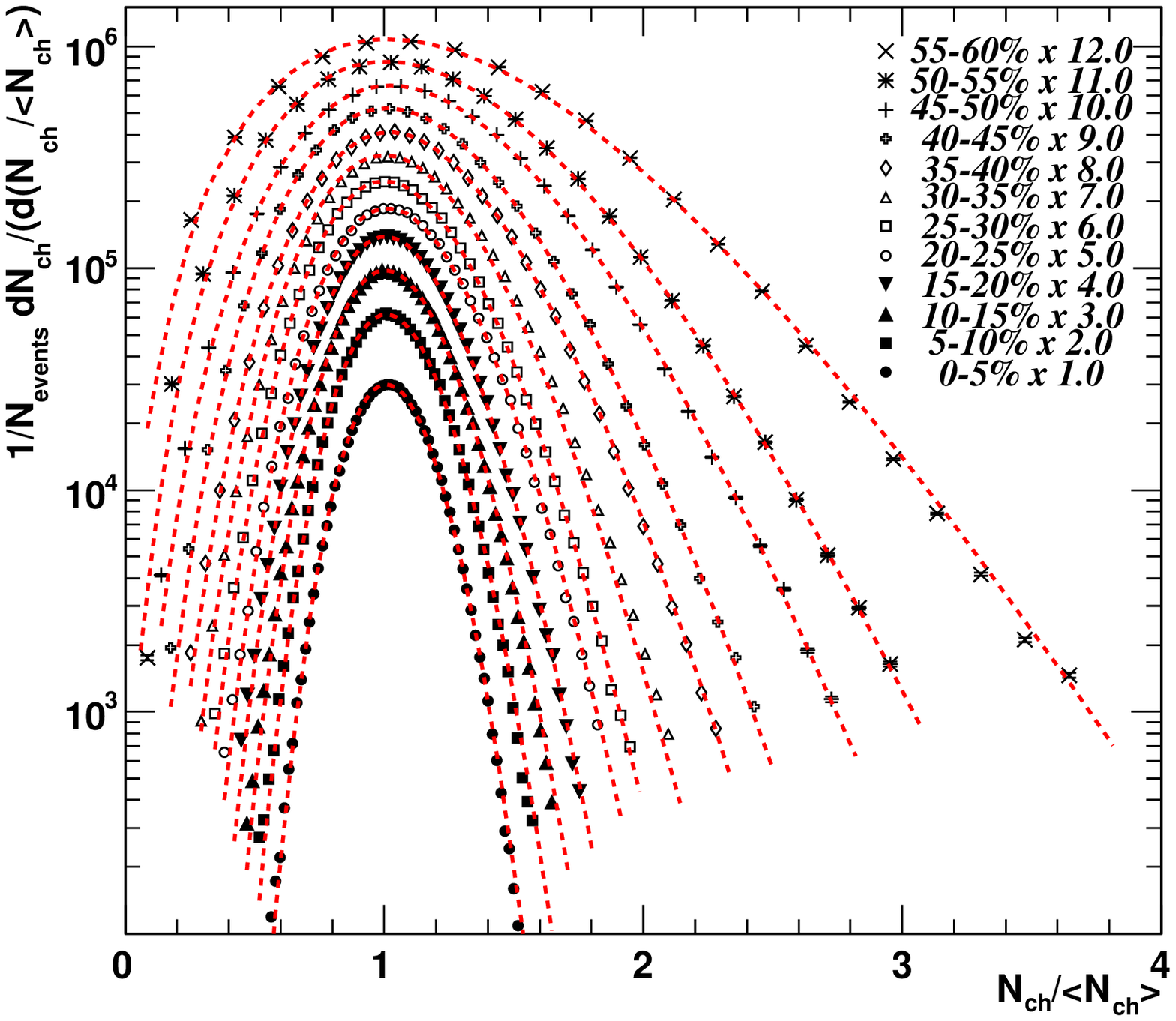} 
\includegraphics[width=1.0\linewidth]{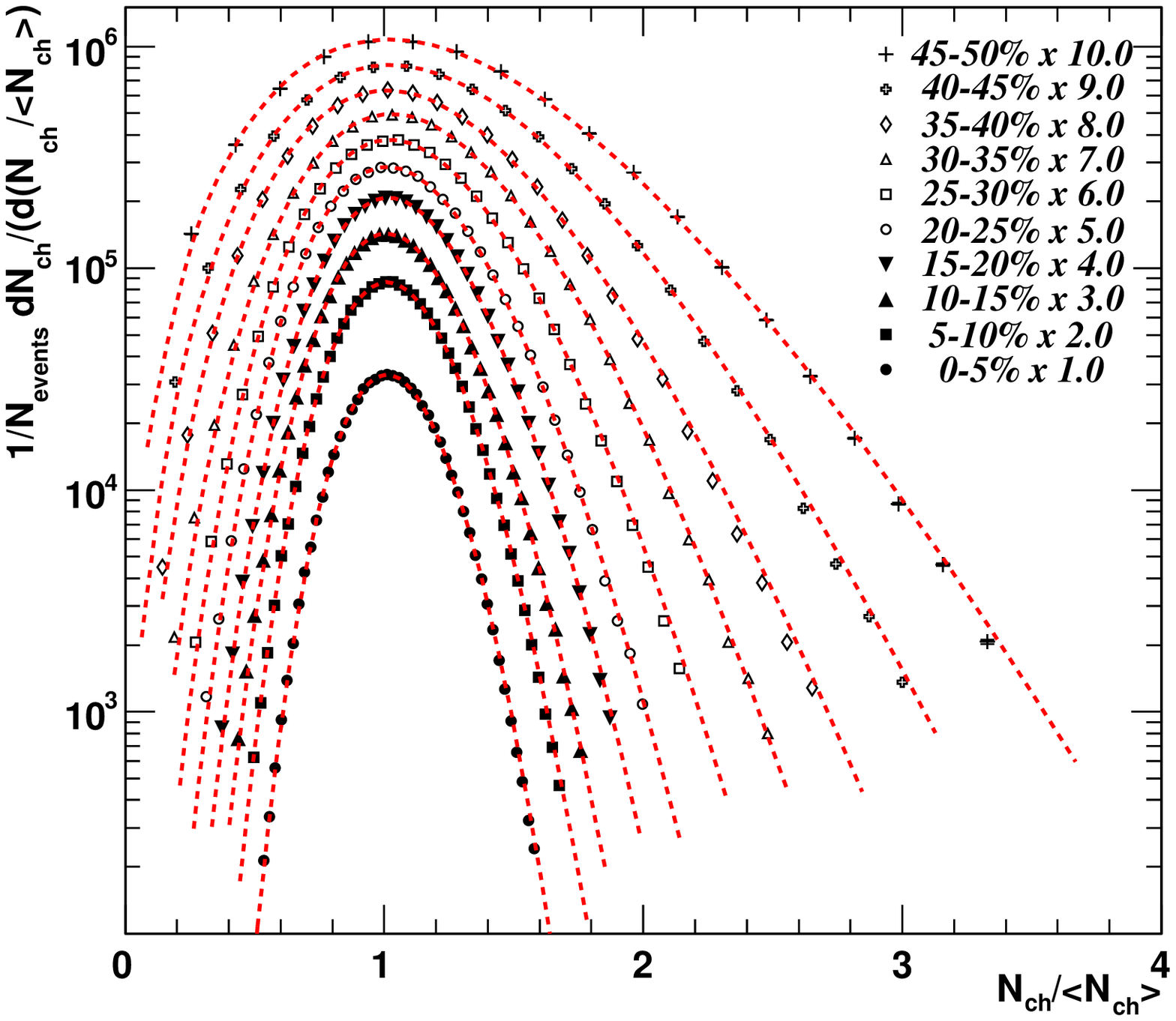} 
\vspace{-1.2cm}
\caption{\label{fig:nbdAuAu}
The uncorrected multiplicity distributions of charged hadrons with 
$0.2<p_T<2.0$ GeV/c for 200 (upper) and 62.4 (lower) GeV Au+Au 
collisions. The dashed lines are fits to the Negative Binomial 
Distribution. The data are normalized to the mean and scaled by the 
amounts in the legend. }
\end{figure}

\begin{figure}[thb]
\vspace{-0.2cm}
\includegraphics[width=1.0\linewidth]{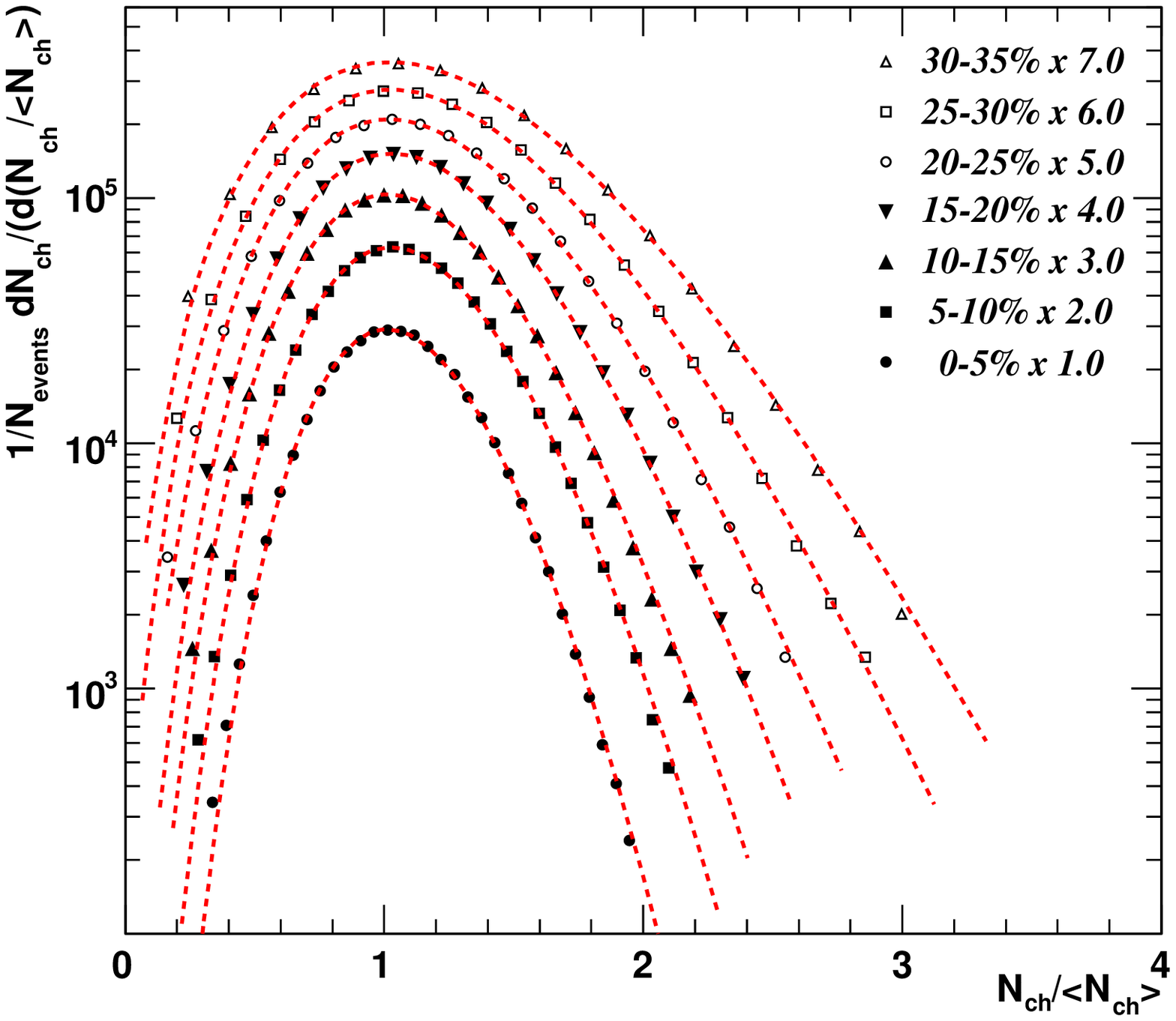} 
\includegraphics[width=1.0\linewidth]{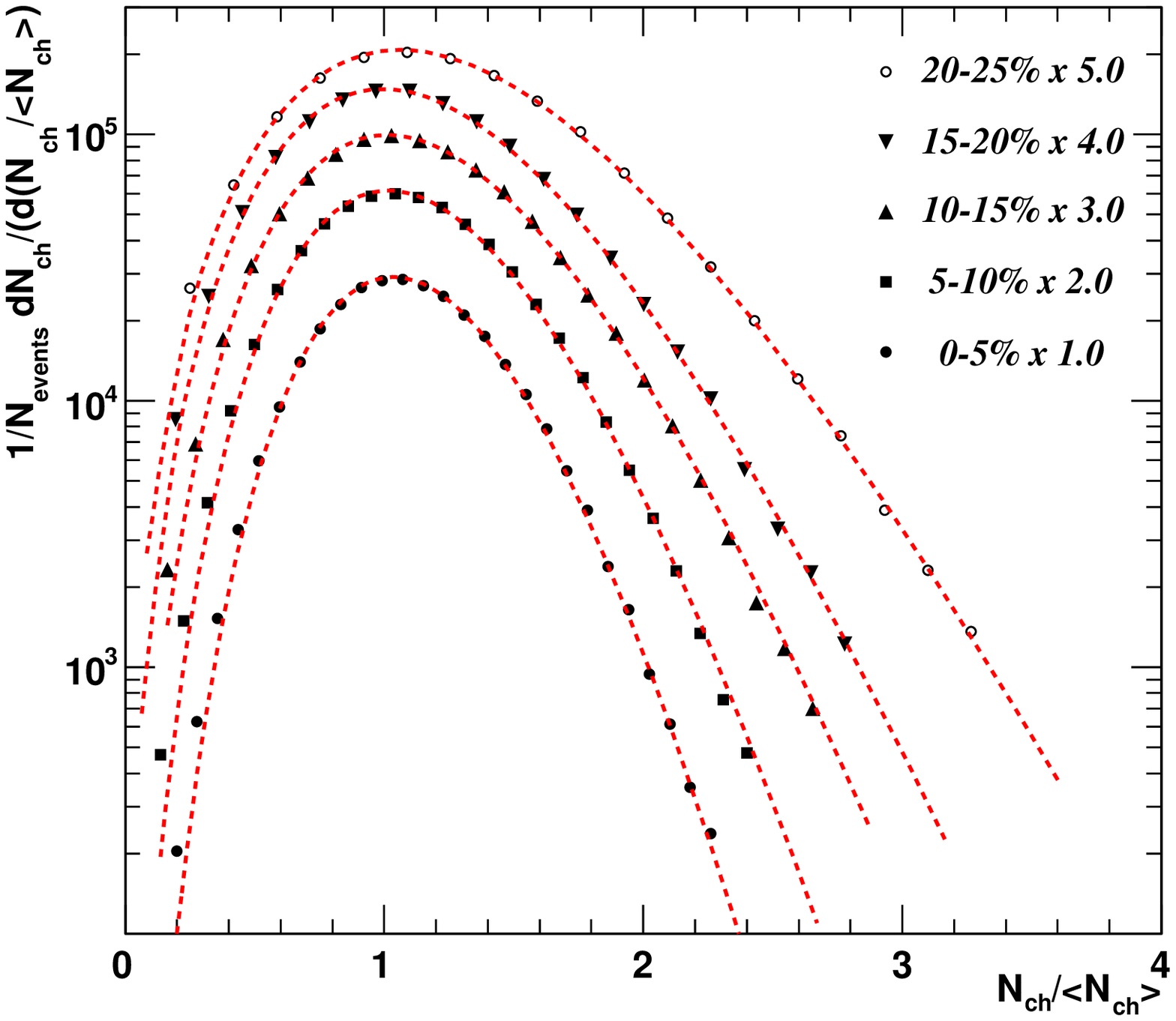} 
\includegraphics[width=1.0\linewidth]{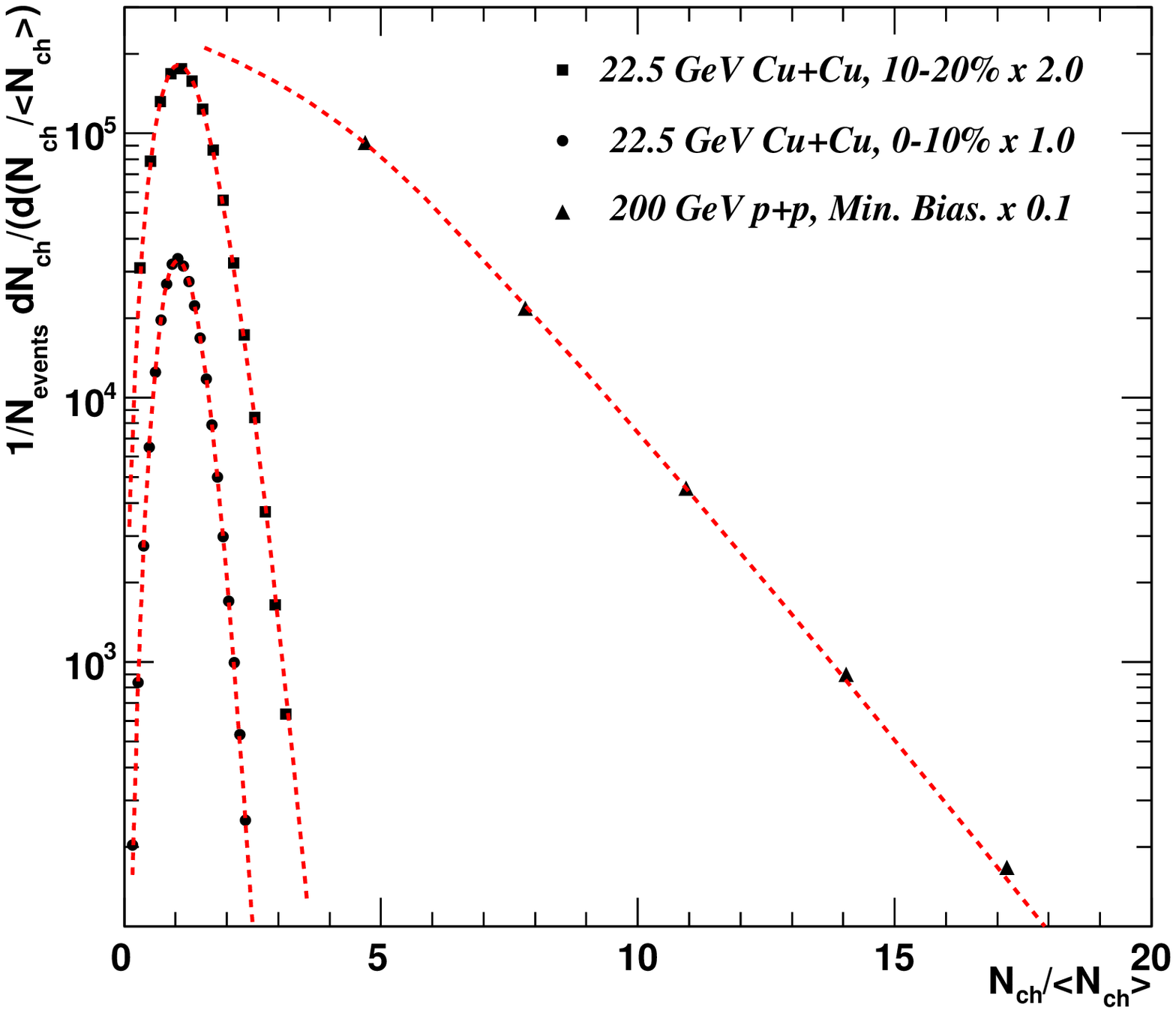} 
\vspace{-1.2cm}
\caption{\label{fig:nbdCuCu}
The uncorrected multiplicity distributions of charged hadrons with 
$0.2<p_T<2.0$ GeV/c for 200 (upper), 62.4 (middle), and 
22.5 (lower) GeV Cu+Cu and 200 GeV p+p (lower) collisions. 
The dashed lines are fits to the Negative Binomial 
Distribution. The data are normalized to the mean and scaled by the 
amounts in the legend. }
\end{figure}

\begin{figure}[th]
\includegraphics[width=0.8\linewidth]{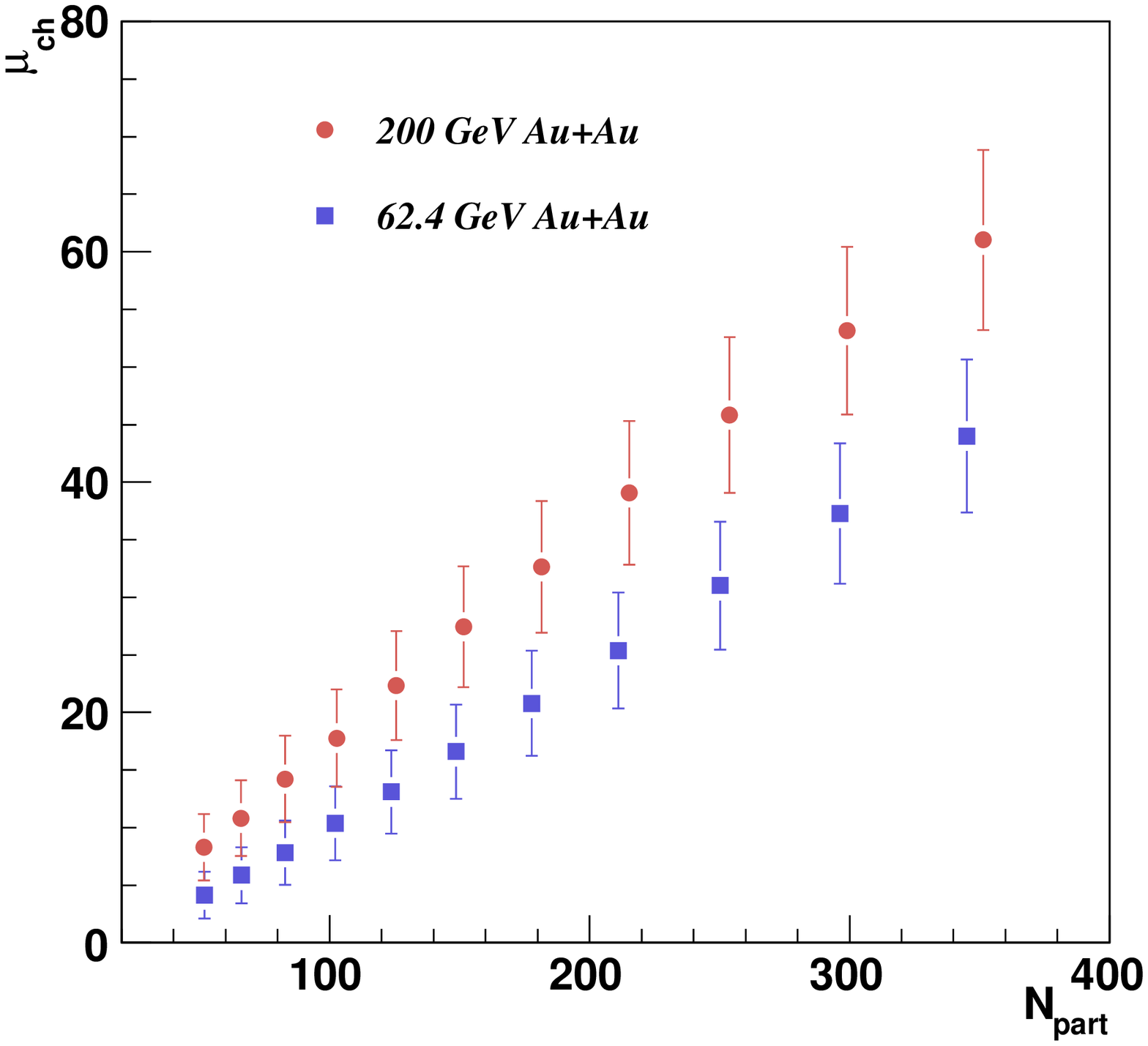} 
\includegraphics[width=1.0\linewidth]{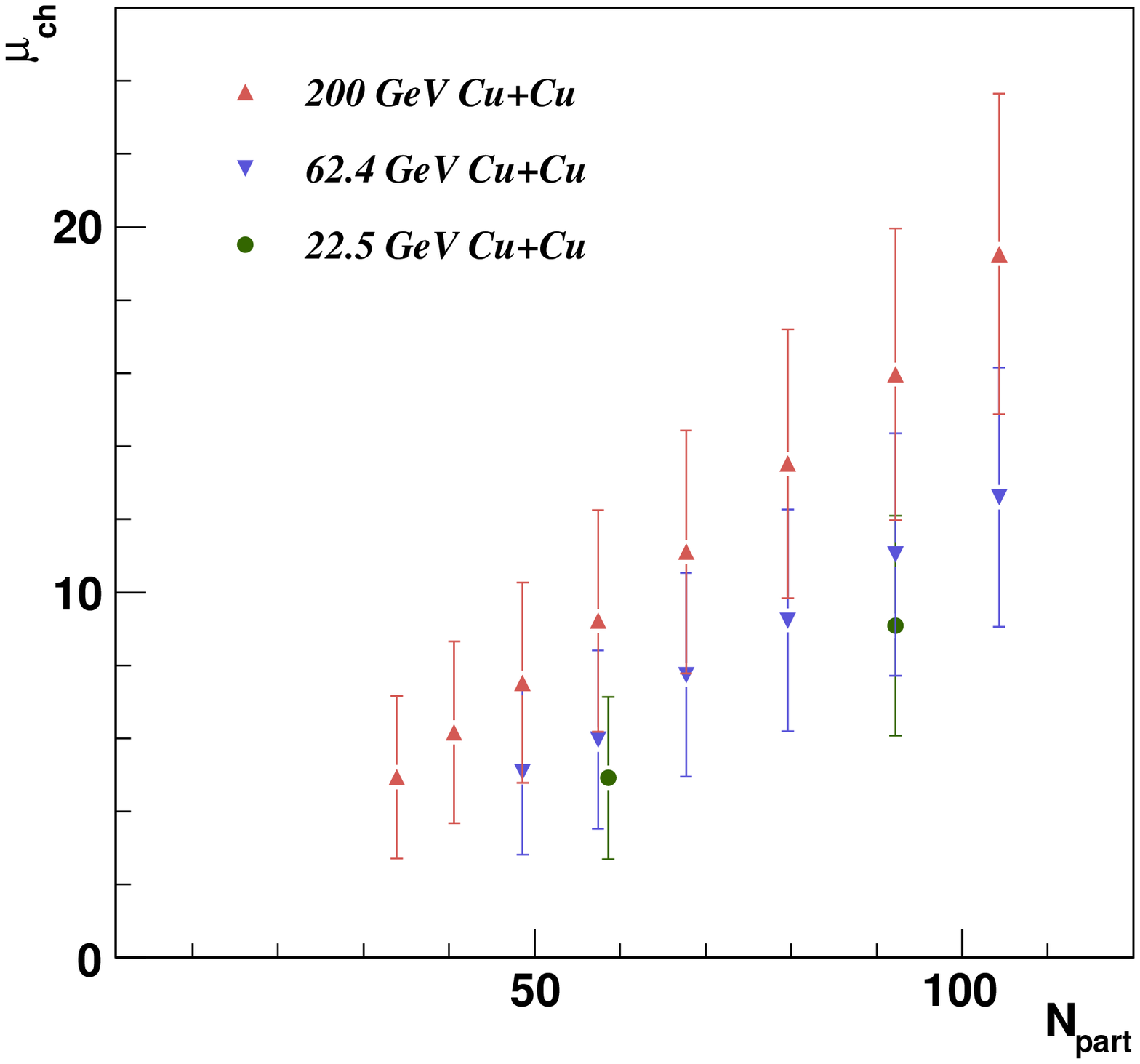} 
\caption{\label{fig:Mean}
The mean from the NBD fit as a function of $N_{\rm part}$ for 
Au+Au (upper) and Cu+Cu (lower)  
collisions over the range $0.2<p_T<2.0$ GeV/c. The mean shown is 
within the PHENIX central arm spectrometer acceptance. The error 
bars represent the standard deviation of the distribution. }
\end{figure}

Figures~\ref{fig:nbdAuAu}-\ref{fig:nbdCuCu} show the 
uncorrected, or raw, multiplicity distributions in the $p_T$ range 
$0.2<p_T<2.0$ GeV for all centralities from each collision system 
overlayed with fits to Negative Binomial Distributions (dashed 
lines). For presentation purposes, the data have been normalized on 
the horizontal axis by the mean of the distribution and scaled on 
the vertical axis by the successive amounts stated in the legend.  
The NBD fits describe the data distributions very well for all 
collision systems, centralities, and $p_T$ ranges. Hence, the mean 
and variance of the multiplicity distributions presented here are 
all extracted from NBD fits. The results of each fit for 
$0.2<p_T<2.0$ GeV are compiled in Table~\ref{tab:Fluc}.
The mean and standard deviation of each fit for $0.2<p_T<2.0$ GeV 
are plotted in Fig.~\ref{fig:Mean}.

Each dataset was taken over spans of several days to several weeks, 
all spanning three separate RHIC running periods. During these 
periods, changes in the total acceptance and efficiency of the 
central arm spectrometers cause the fluctuation measurements to 
vary, thus introducing an additional systematic error to the 
results. This systematic error was minimized by requiring that the 
dataset is stable in quantities that are sensitive to detector 
variations, including the mean charged particle multiplicity, mean 
collision vertex position, and mean centrality.  A time-dependent 
systematic error is applied independently to each point by 
calculating the standard deviation of the scaled variance calculated 
from subsets of the entire dataset, with each subset containing 
about 1 million events. These systematic errors are applied to all 
subsequent results.

The tracking efficiency of the PHENIX central arm spectrometer is 
dependent on centrality, especially in the most central 200 GeV 
Au+Au collisions \cite{phxSpec}. With the assumption that tracking 
inefficiencies randomly effect the multiplicity distribution on an 
event-by-event basis, the effect of inefficiencies on the scaled 
variance can be estimated using Equation (\ref{eq:acc}) where 
$f_{\rm acc}$ is replaced by the inverse of the tracking efficiency, 
1/$f_{\rm eff}$. Tracking efficiency effects the value of the scaled 
variance by 1.5\% at the most. The scaled variance has been 
corrected for tracking inefficiency as a function of centrality for 
all species. The uncertainty of the tracking efficiency estimate is 
typically 2\% and has been propagated into the systematic error 
estimate on a point-by-point basis.

Due to the non-zero width of the centrality bin selection from the 
data, each centrality bin necessarily selects a range of impact 
parameters. This introduces a non-dynamical fluctuation component to 
the measured multiplicity fluctuations due to the resulting 
fluctuations in the geometry of the collisions \cite{ppg061,Kon06}. 
Therefore, it is necessary to estimate the magnitude of the geometry 
fluctuation component so that only the interesting dynamical 
fluctuations remain. The most practical method for estimating the 
geometry fluctuation component is with a model of heavy ion 
collisions. The URQMD \cite{mfURQMD} and HSD \cite{mfHSD,mfHSDURQMD} 
models have previously been applied for this purpose. Here, the 
HIJING event generator \cite{HIJING} is chosen for this estimate 
because it well reproduces the mean multiplicity in heavy ion 
collisions \cite{phxMult} as measured by the PHENIX detector. HIJING 
includes multiple minijet production based upon QCD-inspired models, 
soft excitation, nuclear shadowing of parton distribution functions, 
and the interaction of jets in dense nuclear matter. The estimate is 
performed individually for each centrality bin, collision system, 
and $p_T$ range using the following 
procedure. First, HIJING is run 
with an impact parameter distribution that is sampled from a 
Gaussian distribution with a mean and standard deviation that, for a 
given centrality bin, reproduces the distributions of the charge 
deposited in the BBC and the energy deposited in the ZDC (for 200 
GeV Au+Au).  Second, HIJING is run at a fixed impact parameter with 
a value identical to the mean of the Gaussian distribution in the 
first run.  For each centrality bin, 12,000 HIJING events are 
processed for each impact parameter selection. The scaled variance 
for each impact parameter selection, $\omega_{\rm Gauss}$ and 
$\omega_{\rm fixed}$, is extracted and the measured scaled variance 
is corrected 


\begin{table*}[tbh]   
\caption{\label{tab:Fluc}
Tabulation of the charged hadron multiplicity data and corrections 
for $0.2<p_T<2.0$ GeV/c. The 
errors quoted for $\mu_{\rm ch}$ and $\sigma_{\rm ch}$ represent 
their time-dependent systematic error. The errors quoted for 
$\omega_{\rm ch,dyn}$ and $1/k_{\rm NBD,dyn}$ represent their total 
systematic error. For each dataset the first three columns give the species, collision
energy, and geometric correction factor, $f_{\rm geo}$, respectively.
}
\begin{ruledtabular}
\begin{tabular}{ccccccccc} 
Species & $\sqrt{s_{\rm NN}}$ & $f_{\rm geo}$ & $N_{\rm part}$ & $\mu_{\rm ch}$ & raw $\sigma_{\rm ch}^{2}$ & $\omega_{\rm ch,dyn}$ & $1/k_{\rm NBD,dyn}$ & $\chi^{2}$/dof \\
        & (GeV) & & & & & &  & \\ \hline
&&& 351 & $61.0\pm1.1$ & $75.6\pm1.9$ & $  1.10\pm0.02$ & $1.45\cdot10^{-03}\pm2.2\cdot10^{-04}$ & 37.1/58\\
&&& 299 & $53.1\pm1.0$ & $71.8\pm1.8$ & $  1.15\pm0.02$ & $2.45\cdot10^{-03}\pm2.7\cdot10^{-04}$ & 38.6/56\\
&&& 253 & $45.8\pm0.8$ & $ 65.2\pm1.5$ & $  1.17\pm0.02$ & $3.41\cdot10^{-03}\pm2.9\cdot10^{-04}$ & 34.0/54\\
&&& 215 & $39.1\pm0.7$ & $ 57.8\pm1.6$ & $  1.19\pm0.03$ & $4.53\cdot10^{-03}\pm3.6\cdot10^{-04}$ & 29.1/53\\
&&& 181 & $32.6\pm0.6$ & $ 49.7\pm1.3$ & $  1.21\pm0.03$ & $5.95\cdot10^{-03}\pm5.1\cdot10^{-04}$ & 24.5/50\\
Au+Au & 200 & $0.37\pm0.027$ & 151 & $27.4\pm0.5$ & $ 41.4\pm1.0$ & $ 1.20\pm0.03$ & $6.86\cdot10^{-03}\pm5.5\cdot10^{-04}$ & 20.7/46\\
&&& 125 & $22.3\pm0.4$ & $ 33.8\pm0.9$ & $  1.20\pm0.03$ & $8.47\cdot10^{-03}\pm7.1\cdot10^{-04}$ & 11.9/41\\
&&& 102 & $17.8\pm0.3$ & $ 26.7\pm0.6$ & $  1.19\pm0.02$ & $1.05\cdot10^{-02}\pm9.0\cdot10^{-04}$ & 16.6/37\\
&&&  82 & $14.2\pm0.3$ &  $20.8\pm0.6$ & $  1.17\pm0.02$ & $1.20\cdot10^{-02}\pm1.0\cdot10^{-03}$ & 37.8/33\\
&&&  65 & $10.8\pm0.2$ &  $16.0\pm0.4$ & $  1.18\pm0.02$ & $1.64\cdot10^{-02}\pm1.3\cdot10^{-03}$ & 37.8/28\\
&&&  51 &  $8.3\pm0.2$ &  $12.1\pm0.3$ & $  1.17\pm0.02$ & $2.06\cdot10^{-02}\pm2.0\cdot10^{-03}$ & 53.8/24\\ \hline
&&&  345 & $44.0\pm0.3$ &  $53.6\pm0.5$ & $  1.08\pm  0.02$ & $1.63\cdot10^{-03}\pm2.0\cdot10^{-04}$ & 14.6/54\\
&&&  296 & $37.3\pm0.2$ &  $48.3\pm0.3$ & $  1.11\pm  0.02$ & $2.63\cdot10^{-03}\pm2.6\cdot10^{-04}$ & 13.8/53\\
&&&  250 & $31.0\pm0.2$ &  $39.8\pm0.4$ & $  1.10\pm  0.02$ & $3.00\cdot10^{-03}\pm3.0\cdot10^{-04}$ & 14.0/50\\
&&&  211 & $25.4\pm0.2$ &  $33.6\pm0.5$ & $  1.12\pm  0.02$ & $4.21\cdot10^{-03}\pm4.4\cdot10^{-04}$ & 8.36/44\\
&&&  177 & $20.8\pm0.1$ &  $27.8\pm0.2$ & $  1.12\pm  0.02$ & $5.34\cdot10^{-03}\pm5.5\cdot10^{-04}$ & 19.2/40\\
Au+Au & 62.4 & $0.33\pm0.031$ & 148 & $16.6\pm0.1$ &  $22.8\pm0.3$ & $  1.13\pm  0.02$ & $7.43\cdot10^{-03}\pm7.8\cdot10^{-04}$ & 25.9/37\\
&&&  123 & $13.1\pm0.1$ &  $18.1\pm0.2$ & $  1.13\pm  0.02$ & $9.61\cdot10^{-03}\pm9.7\cdot10^{-04}$ & 34.3/33\\
&&&  102 & $10.4\pm0.1$ &  $14.9\pm0.1$ & $  1.15\pm  0.02$ & $1.38\cdot10^{-02}\pm1.4\cdot10^{-03}$ & 44.5/28\\
&&&  82 &  $7.8\pm0.1$ &  $11.1\pm0.1$ & $  1.14\pm  0.02$ & $1.76\cdot10^{-02}\pm1.9\cdot10^{-03}$ & 50.9/24\\
&&&  66 &  $5.9\pm0.04$ &   $8.3\pm0.1$ & $  1.14\pm  0.02$ & $2.37\cdot10^{-02}\pm3.8\cdot10^{-03}$ & 45.4/20\\
&&&  51 &  $4.1\pm0.03$ &   $5.8\pm0.04$ & $  1.13\pm  0.02$ & $3.08\cdot10^{-02}\pm9.1\cdot10^{-03}$ & 36.2/17\\ \hline
&&&  104 & $19.3\pm0.3$ & $25.7\pm0.8$ & $  1.14\pm  0.03$ & $6.93\cdot10^{-03}\pm1.3\cdot10^{-03}$ & 24.3/30\\
&&&  92 & $16.0\pm0.2$ &  $21.9\pm0.5$ & $  1.15\pm  0.03$ & $9.26\cdot10^{-03}\pm1.5\cdot10^{-03}$ & 21.7/31\\
&&&  79 & $13.5\pm0.2$ &  $18.8\pm0.4$ & $  1.16\pm  0.03$ & $1.15\cdot10^{-02}\pm2.1\cdot10^{-03}$ & 19.4/29\\
Cu+Cu & 200 & $0.40\pm0.047$ & 67 & $11.1\pm0.2$ &  $15.3\pm0.3$ & $  1.15\pm  0.03$ & $1.36\cdot10^{-02}\pm2.0\cdot10^{-03}$ & 29.9/26\\
&&&  57 & $ 9.2\pm0.1$ &  $13.0\pm0.3$ & $  1.17\pm  0.03$ & $1.75\cdot10^{-02}\pm2.5\cdot10^{-03}$ & 26.0/25\\
&&&  48 & $ 7.5\pm0.1$ &  $10.5\pm0.2$ & $  1.16\pm  0.03$ & $2.14\cdot10^{-02}\pm3.6\cdot10^{-03}$ & 30.6/22\\
&&&  40 & $ 6.2\pm0.1$ &  $ 8.7\pm0.2$ & $  1.17\pm  0.03$ & $2.69\cdot10^{-02}\pm4.8\cdot10^{-03}$ & 28.6/20\\
&&&  33 & $ 4.9\pm0.06$ &  $ 6.8\pm0.1$ & $  1.16\pm  0.03$ & $3.12\cdot10^{-02}\pm8.5\cdot10^{-03}$ & 45.7/18\\ \hline
&&&  104 & $12.6\pm0.1$ &  $16.7\pm0.2$ & $  1.10\pm  0.03$ & $8.16\cdot10^{-03}\pm1.7\cdot10^{-03}$ & 40.6/31\\
&&&  92 & $11.0\pm0.1$ &  $16.2\pm0.1$ & $  1.15\pm  0.04$ & $1.35\cdot10^{-02}\pm2.7\cdot10^{-03}$ & 64.2/30\\
Cu+Cu & 62.4 & $0.32\pm0.063$ & 79 &  $9.2\pm0.1$ &  $14.3\pm0.2$ & $  1.18\pm  0.05$ & $1.92\cdot10^{-02}\pm3.9\cdot10^{-03}$ & 37.0/28\\
&&&  67 &  $7.7\pm0.1$ &  $12.0\pm0.2$ & $  1.18\pm  0.05$ & $2.29\cdot10^{-02}\pm4.6\cdot10^{-03}$ & 32.0/26\\
&&&  57 &  $6.0\pm0.1$ &  $ 9.1\pm0.1$ & $  1.17\pm  0.05$ & $2.85\cdot10^{-02}\pm5.9\cdot10^{-03}$ & 32.0/23\\
&&&  48 &  $5.1\pm0.1$ &  $ 8.1\pm0.1$ & $  1.19\pm  0.05$ & $3.66\cdot10^{-02}\pm8.0\cdot10^{-03}$ & 29.2/21\\ \hline
&&& 92 &  $9.1\pm0.04$ &  $10.3\pm0.1$ & $  1.04\pm  0.02$ & $4.31\cdot10^{-03}\pm9.8\cdot10^{-04}$ & 7.45/24\\
Cu+Cu & 22.5 & $0.30\pm0.064$ & 58 &  $4.9\pm0.02$ &  $ 5.8\pm0.04$ & $  1.06\pm  0.02$ & $1.11\cdot10^{-02}\pm2.9\cdot10^{-03}$ & 71.1/17\\
\end{tabular}
\end{ruledtabular}
\end{table*}

\clearpage


\noindent as the fractional deviation from a scaled variance of 
1.0 of a Poisson distribution as follows:
\begin{equation} \label{eq:bGeo}
  \omega_{\rm ch,dyn} - 1 = \frac{(\omega_{\rm fixed}-1)}{(\omega_{\rm Gauss}-1)}~(\omega_{\rm ch,raw}-1) = f_{\rm geo} (\omega_{\rm ch,raw}-1),
\end{equation}
where $\omega_{\rm ch,dyn}$ represents the estimate of the remaining 
dynamical multiplicity fluctuations and $\omega_{\rm ch,raw}$ 
represents the uncorrected multiplicity fluctuations. Since the 
correction, $f_{\rm geo}$, is calculated as a ratio of the two 
running conditions of the simulation, most multiplicity fluctuations 
intrinsic to the model should be canceled. The correction always 
reduces the magnitude of the measured scaled variance. Note that the 
value of $f_{\rm geo}$ is mathematically identical when applied to 
the inverse of $k_{\rm NBD}$:
\begin{equation} \label{eq:kGeo}
  k_{\rm NBD,dyn}^{-1} = f_{\rm geo} k_{\rm NBD}^{-1}.
\end{equation}

\begin{figure}[t]
\includegraphics[width=1.0\linewidth]{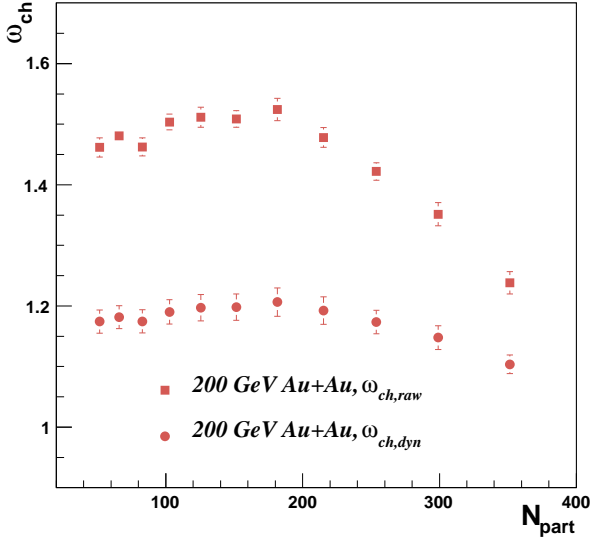} 
\caption{\label{fig:auau200SvarRawCorr}
Fluctuations expressed as the scaled variance as a function of 
centrality for 200 GeV Au+Au collisions in the range $0.2<p_T<2.0$ 
GeV/c. Shown are the uncorrected fluctuations, $\omega_{\rm 
ch,raw}$, along with fluctuations after correcting for the estimated 
contribution from geometry fluctuations using Equation 
\ref{eq:bGeo}, $\omega_{\rm ch,dyn}$. }
\end{figure}

The resulting geometrical correction factors for each species are 
constant as a function of centrality, therefore a single correction 
factor is calculated for each transverse momentum range by fitting 
the correction factors as a function of $N_{\rm part}$ to a 
constant.  This behavior is expected since centrality bins are 
defined to be constant percentages of the total geometric cross 
section. The correction factors for each transverse momentum range 
for a given collision species are consistent with each other. The 
standard deviation of the individual geometrical correction factors 
from the linear fits as a function of $N_{\rm part}$ are included in 
the systematic error of the correction factor estimation and 
propagated into the total systematic error for each point in 
$\omega_{\rm ch,dyn}$ and $k_{\rm NBD,dyn}$.  For $0.2<p_T<2.0$ 
GeV/c, the geometrical correction factors, $f_{\rm geo}$, and 
systematic errors from the fit are $0.37\pm0.027$ for 200 GeV Au+Au, 
$0.33\pm0.031$ for 62.4 GeV Au+Au, $0.40\pm0.047$ for 200 GeV Cu+Cu, 
$0.32\pm0.063$ for 62.4 GeV Cu+Cu, and $0.30\pm0.064$ for 22.5 GeV 
Cu+Cu. The extraction of the geometrical correction factors are 
inherently model-dependent and are also dependent on the accuracy 
with which the centrality detectors are modelled.  The effect of the 
latter dependence has been studied by also calculating the 
correction factors using constant but non-overlapping impact 
parameter distributions for each centrality bin and comparing them 
to the correction factors using the Gaussian impact parameter 
distributions. For all $p_T$ ranges, an additional fraction of the 
value of $\omega_{\rm ch,dyn}$ or $k_{\rm NBD,dyn}^{-1}$ has been 
included in the final systematic errors for these quantities. The 
magnitude of this systematic error is 8\% for 200 GeV Au+Au, 8\% for 
62.4 GeV Au+Au, 11\% for 200 GeV Cu+Cu, 17\% for 62.4 GeV Cu+Cu, and 
25\% for 22.5 GeV Cu+Cu. A sample comparison of the scaled variance 
before and after the application of the geometrical correction 
factor is shown for the 200 GeV Au+Au dataset in 
Fig.~\ref{fig:auau200SvarRawCorr}.

\section{Results}

\begin{figure}[b]
\includegraphics[width=1.0\linewidth]{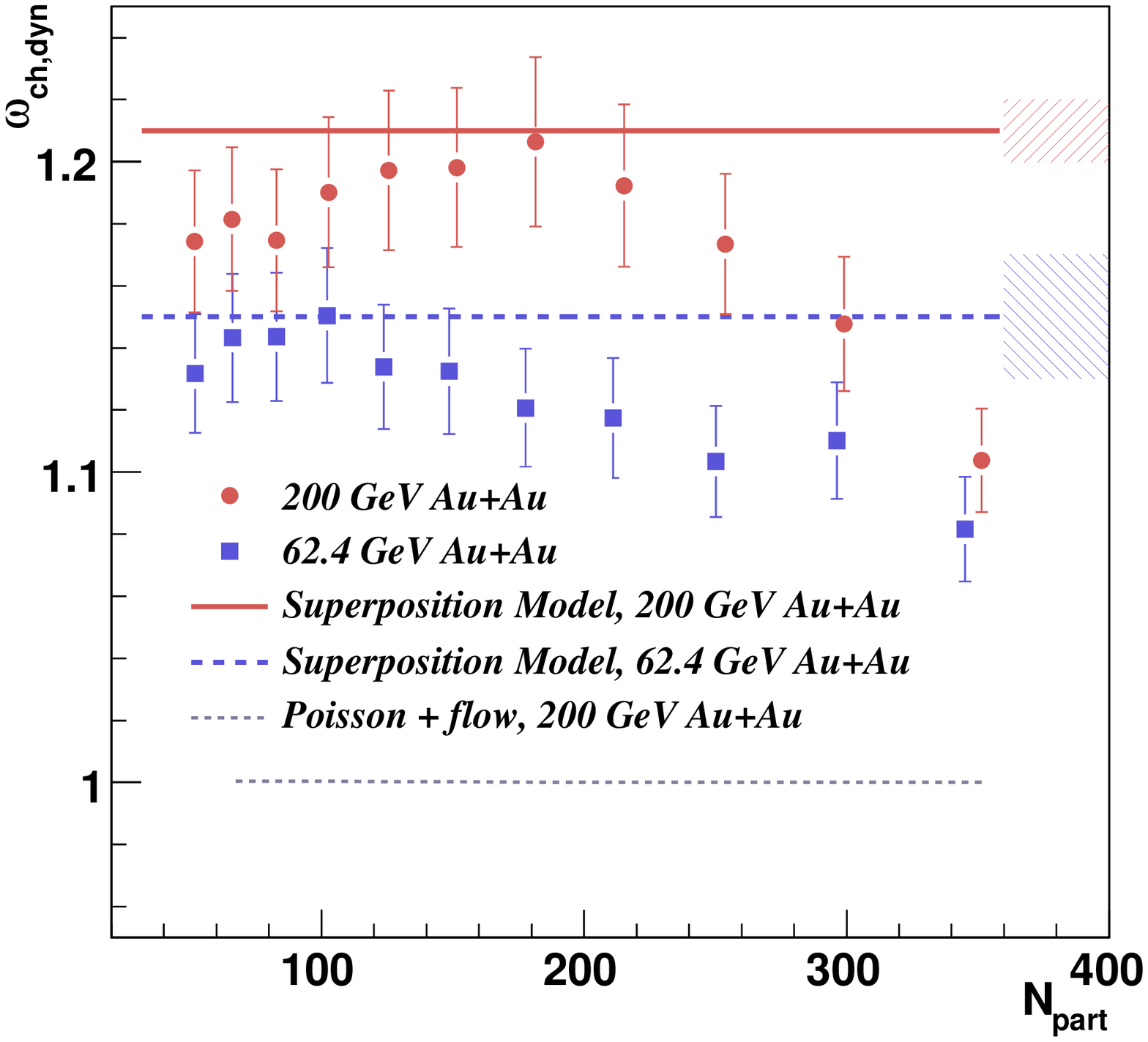} 
\includegraphics[width=1.0\linewidth]{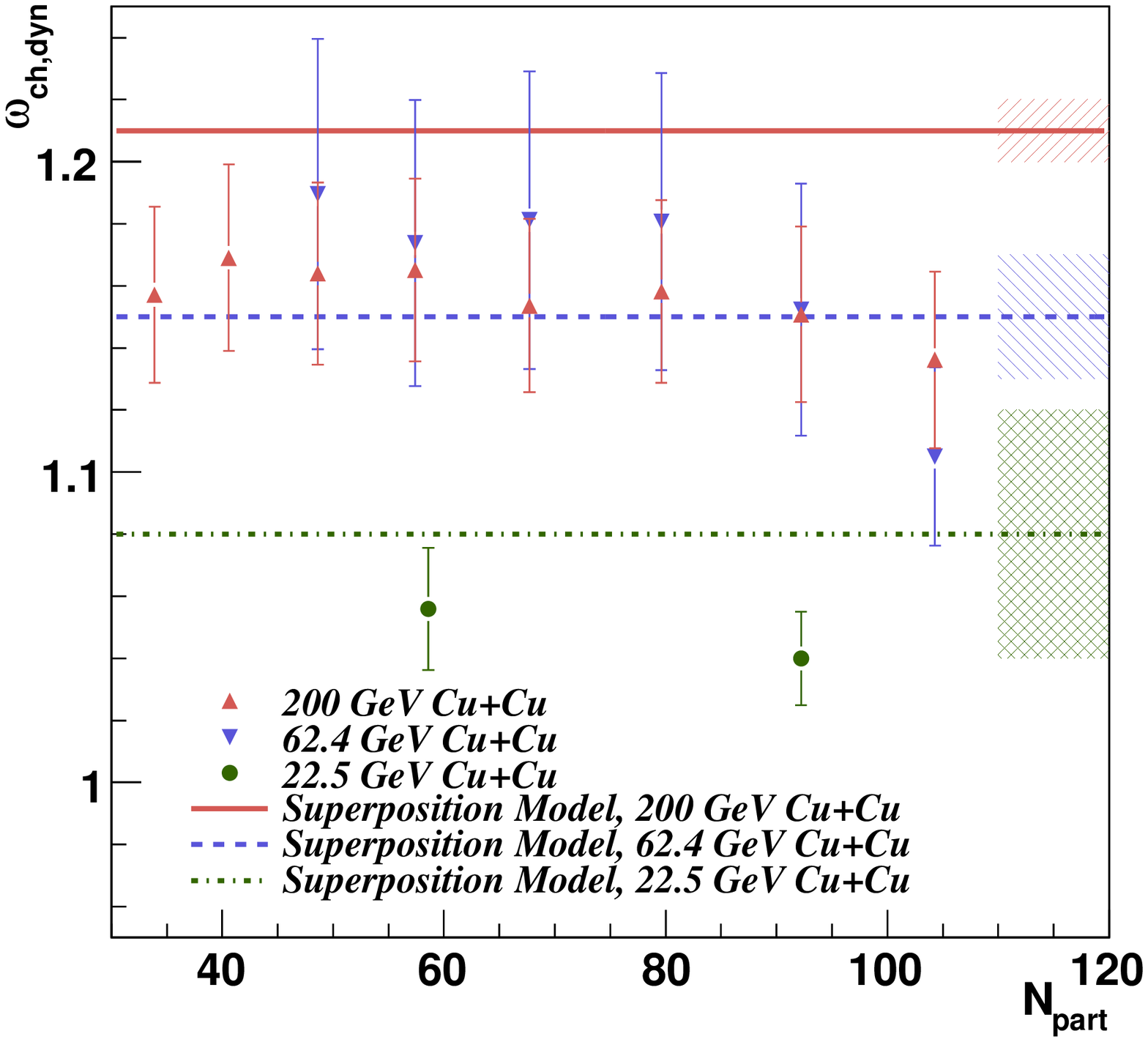} 
\caption{\label{fig:Svar}
Fluctuations expressed as the scaled variance as a function of 
$N_{\rm part}$ for Au+Au (upper) and Cu+Cu (lower) collisions 
for $0.2<p_T<2.0$ GeV/c. The estimated contribution from geometry 
fluctuations has been removed. 
Results from the superposition model are overlayed 
with the shaded regions representing a one standard deviation range 
of the prediction for the fluctuation magnitude derived from p+p 
collision data. Also shown (upper) is the estimated contribution from 
non-correlated particle emission with the Poisson distribution of 
the scaled variance of 1.0 with the addition of elliptic flow in 200 
GeV Au+Au collisions. }
\end{figure}

The scaled variance as a function of the number of participating 
nucleons, $N_{\rm part}$, over the $p_T$ range $0.2<p_T<2.0$ GeV/c 
is shown in Fig.~\ref{fig:Svar}.  For all 
centralities, the scaled variance values consistently lie above the 
Poisson distribution value of 1.0. In all collision systems, the 
minimum scaled variance occurs in the most central collisions and 
then begins to increase as the centrality decreases. In 200 GeV 
Au+Au collisions, this increase is only observed for $N_{\rm 
part}>200$.  For $N_{\rm part}<200$ $\omega_{\rm ch,dyn}$ suggests a 
slight decrease but is consistent with a constant value.  In 62.4 
GeV Au+Au collisions, the increase in $\omega_{\rm ch,dyn}$ with 
decreasing centrality is observed only over the range $N_{\rm 
part}>110$. The source of the qualitative differences between the 
200 and 62.4 GeV Au+Au collisions is not known, although some of the 
differences could be explained by the increased contribution from 
hard scattering processes at 200 GeV compared to 62.4 GeV. Studies 
performed by varying the centrality selection cuts establish that 
the differences are not due to the differences in the centrality 
selection algorithm.  A similar centrality-dependent trend of the 
scaled variance has also been observed at the SPS in low energy 
Pb+Pb collisions at $\sqrt{s_{\rm NN}}$=17.3 GeV and at forward 
rapidities (1.1$<y_{\rm c.m.}<$2.6), measured by experiment NA49 
\cite{na49MF}, where the hard scattering contribution is expected to 
be small. The Cu+Cu data exhibit a weaker decrease in the scaled 
variance for more central collisions. The 62.4 GeV Cu+Cu scaled 
variance values are consistently above those from the 200 GeV Cu+Cu 
dataset, but the two are consistent within the systematic errors for 
all centralities.

\begin{figure}[thb]
\includegraphics[width=1.0\linewidth]{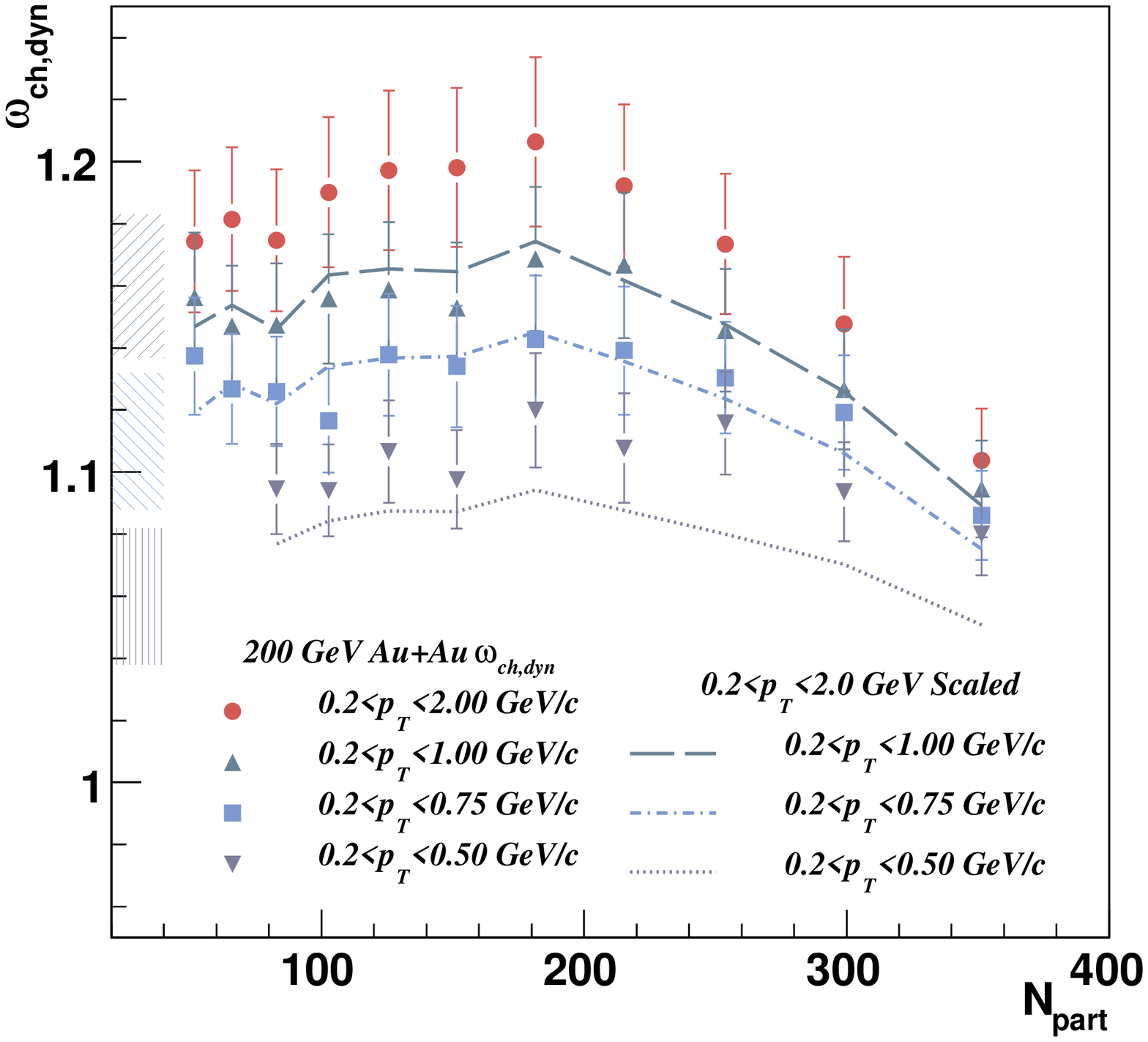} 
\includegraphics[width=1.0\linewidth]{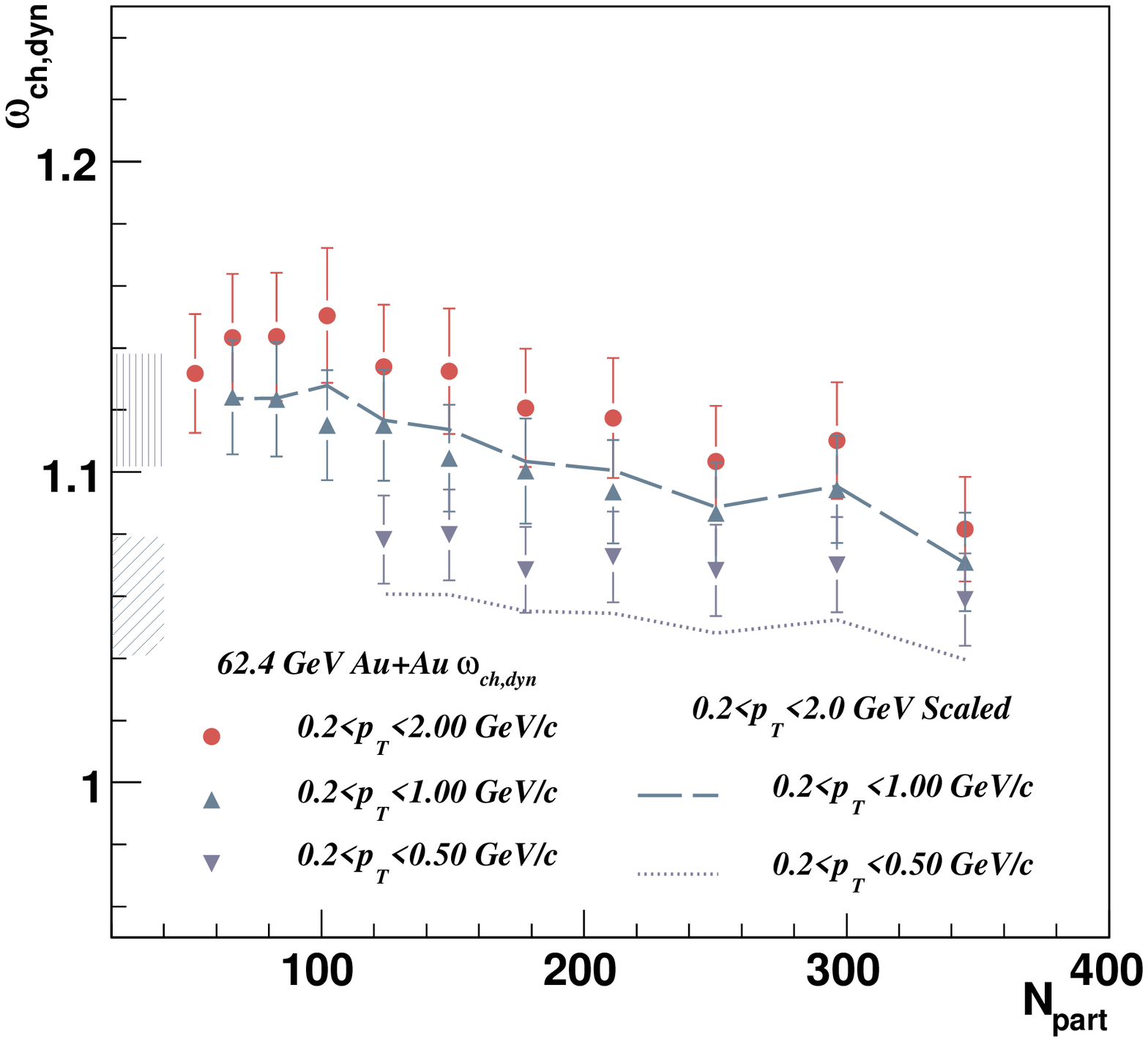} 
\caption{\label{fig:auauSvarPt}
Scaled variance for 200 (upper) and 62.4 (lower) GeV 
Au+Au collisions plotted as a function 
of $N_{\rm part}$ for several $p_T$ ranges. The lines represent the 
data for the reference range $0.2<p_T<2.0$ scaled down using the 
mean multiplicity in each successive $p_T$ range. The shaded areas 
represent the systematic errors from the reference range. }
\end{figure}

\begin{figure}[thb]
\includegraphics[width=1.0\linewidth]{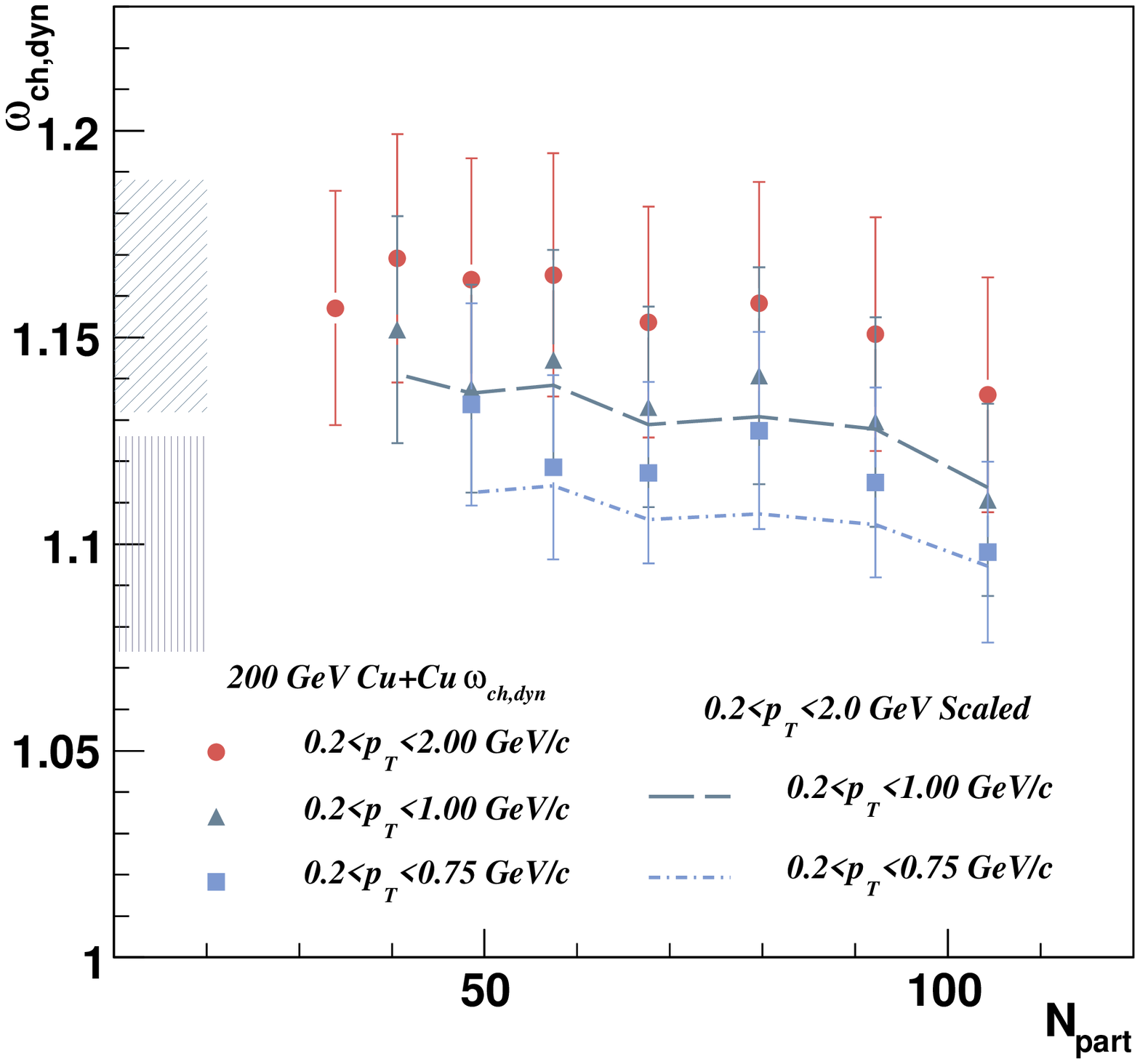} 
\includegraphics[width=1.0\linewidth]{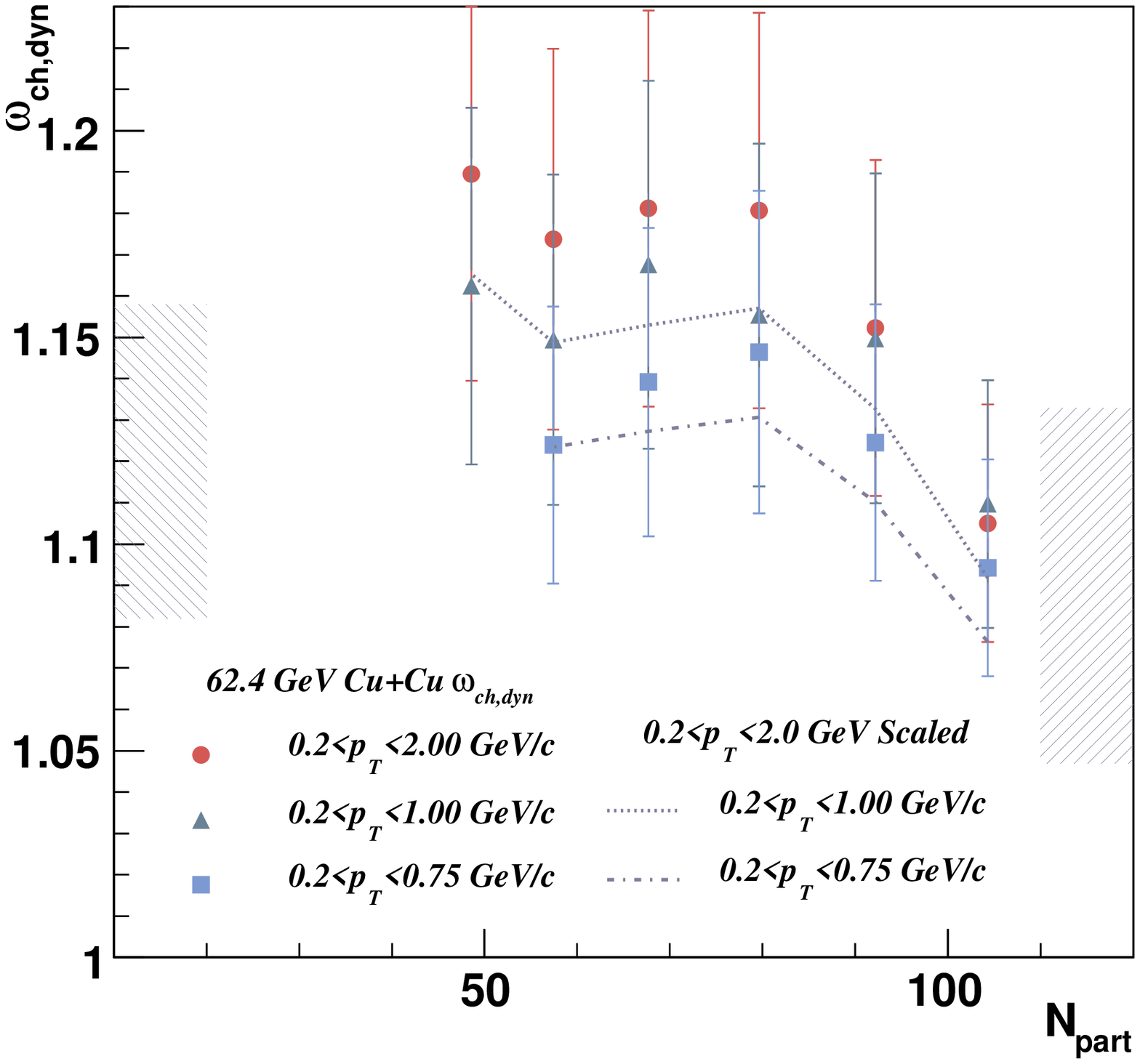} 
\caption{\label{fig:cucuSvarPt}
Scaled variance for 200 (upper) and 62.4 (lower) GeV
Cu+Cu collisions plotted as a function 
of $N_{\rm part}$ for several $p_T$ ranges. The lines represent the 
data for the reference range $0.2<p_T<2.0$ scaled down using the 
mean multiplicity in each successive $p_T$ range. The shaded areas 
represent the systematic errors from the reference range. }
\end{figure}

\begin{figure}[thb]
\includegraphics[width=1.0\linewidth]{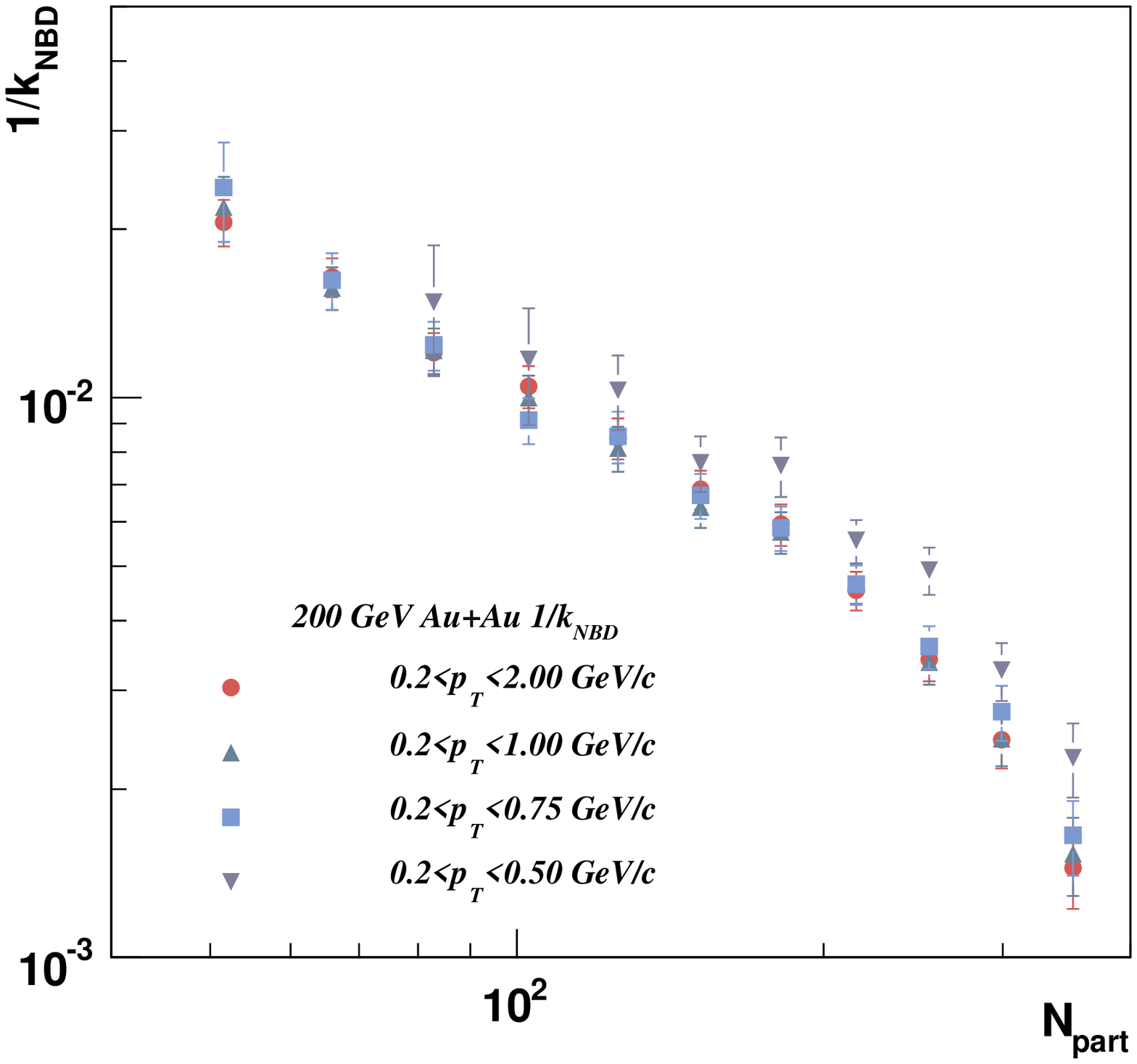} 
\includegraphics[width=1.0\linewidth]{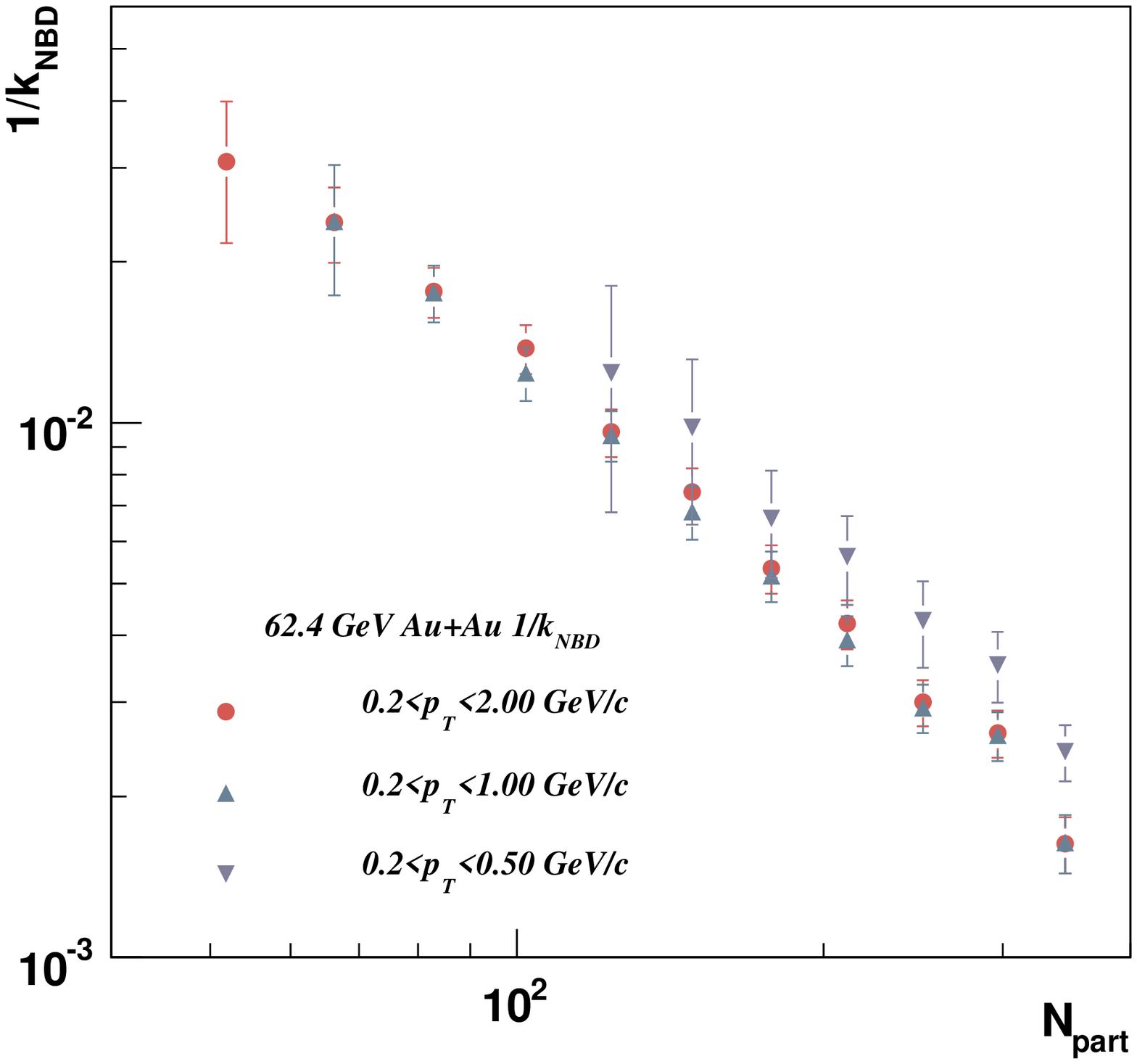} 
\caption{\label{fig:auauInvkPt}
The inverse of the parameter $k_{\rm NBD}$ from the Negative 
Binomial Distribution fits for 200 (upper) and 62.4 (lower) GeV
Au+Au collisions. The 
fluctuations are plotted as a function of $N_{\rm part}$ for several 
$p_T$ ranges. }
\end{figure}

\begin{figure}[thb]
\includegraphics[width=1.0\linewidth]{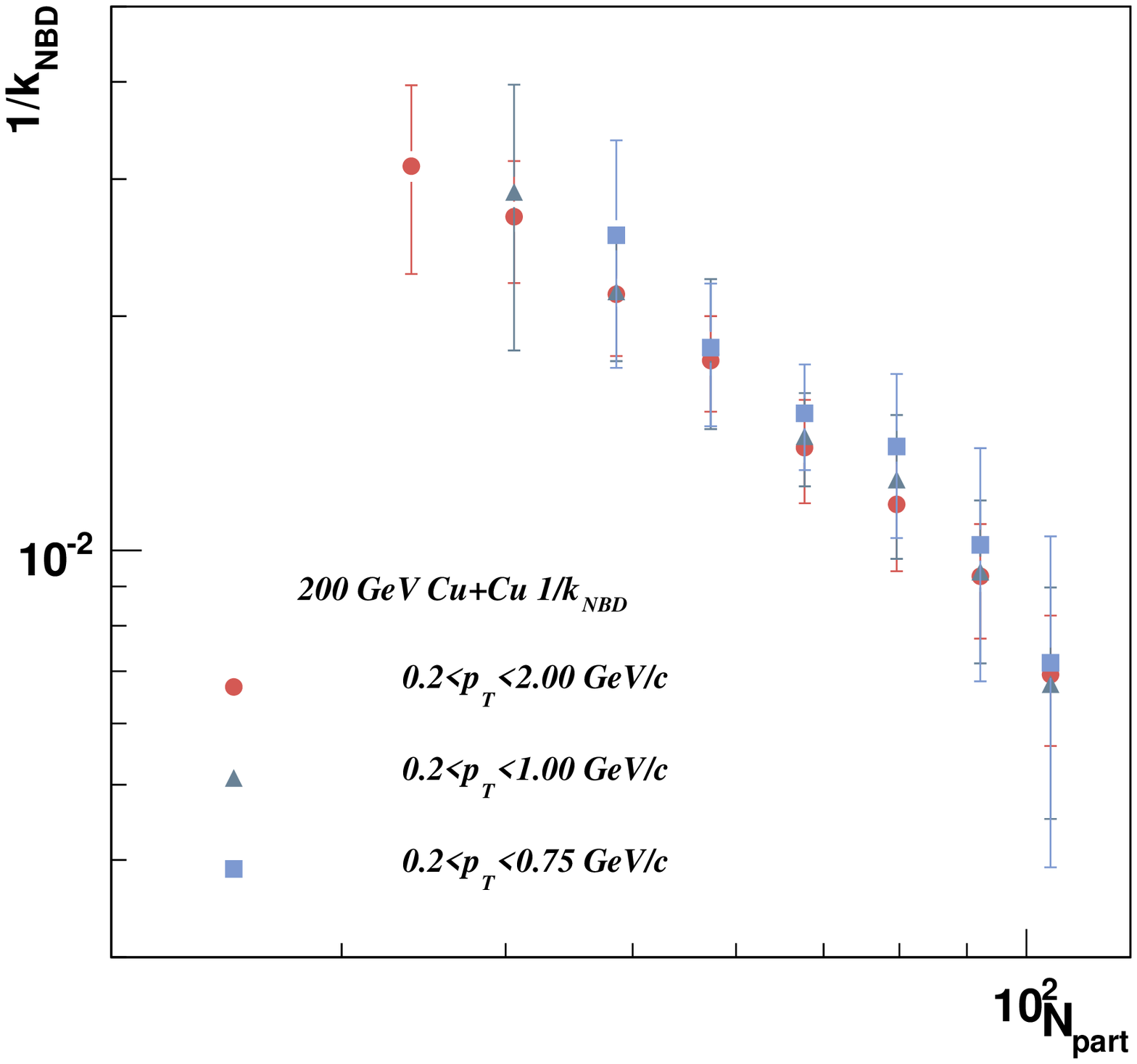} 
\includegraphics[width=1.0\linewidth]{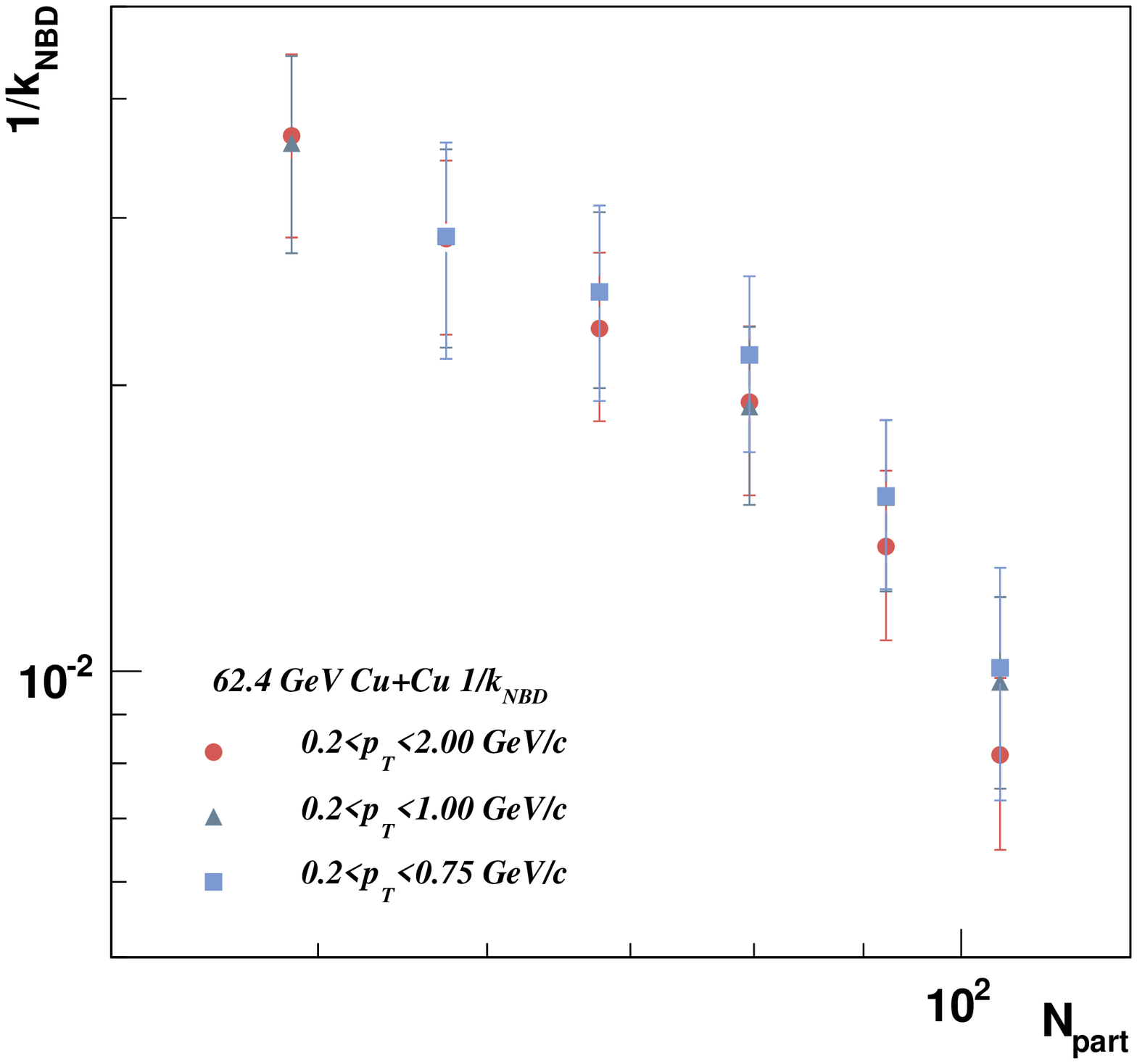} 
\caption{\label{fig:cucuInvkPt}
The inverse of the parameter $k_{\rm NBD}$ from the Negative 
Binomial Distribution fits for 200 (upper) and 62.4 (lower) GeV
Cu+Cu collisions. The 
fluctuations are plotted as a function of $N_{\rm part}$ for several 
$p_T$ ranges. }
\end{figure}

The scaled variance has been studied as a function of the $p_T$ 
range over which the multiplicity distributions are measured in 
order to determine if any significant $p_T$-dependent dynamical 
fluctuations are present. Results for several $p_T$ ranges from 
$0.2<p_T<2.0$ GeV/c down to $0.2<p_T<0.5$ GeV/c are shown in 
Figs.~\ref{fig:auauSvarPt}-\ref{fig:cucuSvarPt}. In the absence of 
$p_T$-dependent dynamical fluctuations, restricting the $p_T$ range 
should reduce the scaled variance in the same manner as for a 
fractional acceptance. Similar to Equation (\ref{eq:acc}):
\begin{equation}  \label{eq:ptacc}
\omega_{\rm p_T} = 1 + f_{\rm pt}(\omega_{\rm ref}-1), 
\end{equation}
where $\omega_{\rm p_T}$ represents the fluctuations in the $p_T$ 
range of interest, $\omega_{\rm ref}$ represents the fluctuations in 
the reference $p_T$ range, and $f_{\rm p_T} = \mu_{\rm p_T}/\mu_{\rm 
ref}$ is the ratio of the mean multiplicity in the two ranges. 
Also shown are curves representing the expected scaling of the 
fluctuations using the range $0.2<p_T<2.0$ GeV/c as the reference 
range. The shaded regions reflect the systematic errors of the 
reference range. For all $p_T$ ranges, the scaled fluctuation curves 
are consistent with the data, indicating that no significant 
$p_T$-dependence is observed, although the data in the range 
$0.2<p_T<0.5$ GeV are consistently above the scaled reference 
curves. The $p_T$-dependence can also be examined in more directly 
with the parameter $k_{\rm NBD}$ from the NBD fits. Substitution of 
the scaled variance in Equation (\ref{eq:omegak}) into Equation 
(\ref{eq:ptacc}) shows that $k_{\rm NBD}$ should be independent of 
$p_T$ in the absence of $p_T$-dependent dynamical fluctuations. As 
shown in Figs.~\ref{fig:auauInvkPt}-\ref{fig:cucuInvkPt}, 
there is no significant $p_T$-dependence of the observed values of 
$k_{\rm NBD}$.



\begin{figure}[thb]
\includegraphics[width=1.0\linewidth]{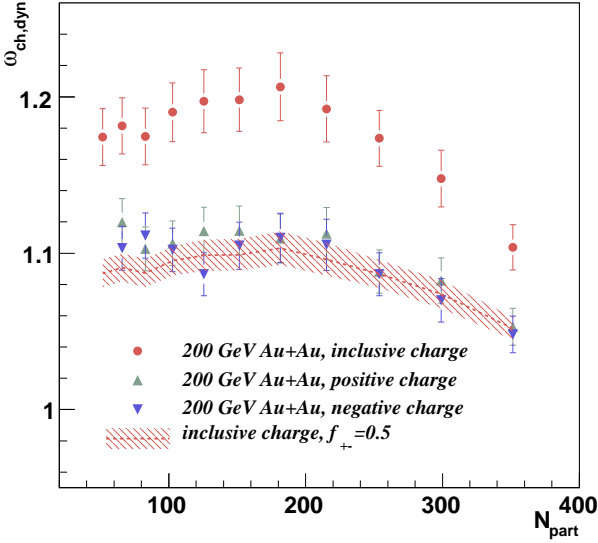} 
\caption{\label{fig:auau200SvarQ}
The scaled variance as a function of $N_{\rm part}$ for 200 GeV 
Au+Au collisions in the range $0.2<p_T<2.0$ GeV/c. Shown are 
measurements for inclusive charged particles, positive particles, 
and negative particles.  The line represents the inclusive data 
scaled down in acceptance by 50\% with the shaded area representing 
the systematic error. }
\end{figure}

The scaled variance as a function of the charge sign of the charged 
hadrons is shown in Fig.~\ref{fig:auau200SvarQ} for 200 GeV Au+Au 
collisions in the $p_T$ range $0.2<p_T<2.0$ GeV/c in order to 
investigate any Coulomb-based contributions to the fluctuations.  
In the absence of additional dynamic fluctuations, the scaled 
variance for positively or negatively charged hadrons should be 
reduced from the inclusive charged hadron value by
\begin{equation} \label{eq:qacc}
  \omega_{\rm +-} = 1 +f_{\rm +-}(\omega_{\rm ch}-1), 
\end{equation}
where $\omega_{\rm +-}$ are the fluctuations for positive or 
negative particles, $\omega_{\rm ch}$ are the fluctuations for 
inclusive charged hadrons, and $f_{\rm +-} = \mu_{\rm +-}/\mu_{\rm 
ch}$ is the ratio of the mean multiplicities. The scaled variance 
from the positive and negative hadrons are consistent with each 
other and consistent with the expected reduction of the inclusive 
charged hadron fluctuations.

An additional non-dynamic contribution to multiplicity fluctuations 
arises from the presence of elliptic flow. This contribution has 
been estimated using a simple Monte Carlo simulation. In this 
simulation, a random reaction plane angle is assigned to each event. 
The multiplicity distribution due to the elliptic flow component is 
given by the following:
\begin{equation} \label{eq:flow}
   dN/d\phi = C[1 + 2~v_2~cos(2\Delta\phi)],
\end{equation}
where C is a normalization factor, $v_2$ is the measured magnitude 
of the elliptic flow, and $\Delta\phi$ is the difference between the 
particle emission angle and the reaction plane angle. For each 
event, this multiplicity distribution function is integrated over 
the PHENIX azimuthal acceptance and the resulting scaled variance 
from one million events is calculated. The value of $v_2$ used in 
the simulation is taken from PHENIX measurements of elliptic flow at 
the mean transverse momentum of the inclusive charged hadron spectra 
in the range $0.2<p_T<2.0$ GeV/c within the central arm 
spectrometers \cite{phenixFlow}. The estimated contribution from 
elliptic flow to the observed scaled variance is less than 0.038\% 
for all centralities and is shown for 200 GeV Au+Au collisions in 
Fig.~\ref{fig:Svar}.

\section{Discussion}

\subsection{Comparisons to a participant superposition model}

It is informative to compare fluctuations in relativistic heavy ion 
collisions to what can be expected from the superposition of 
individual participant nucleon-nucleon collisions. For this purpose, 
PHENIX data will be compared to a participant superposition, or 
wounded nucleon, model \cite{wnm} based upon data from elementary 
collisions. In the participant superposition model, the total 
multiplicity fluctuations can be expressed in terms of the scaled 
variance \cite{heiselReview},
\begin{equation} \label{eq:wnmSvar}
  \omega_{\rm N} = \omega_{\rm \nu} + \mu_{\rm WN}~\omega_{\rm N_{\rm part}}, 
\end{equation}
where $\omega_{\rm \nu}$ are the fluctuations from each individual 
source, e.g. from each elementary collision, $\omega_{\rm N_{\rm 
part}}$ are the fluctuations of the number of sources, and $\mu_{\rm 
WN}$ is the mean multiplicity per wounded nucleon. The second term 
includes non-dynamic contributions from geometry fluctuations due to 
the width of the centrality bin along with additional fluctuations 
in the number of participants for a fixed impact parameter. Ideally, 
the second term is nearly nullified after applying the geometry 
corrections described previously, so the resulting fluctuations are 
independent of centrality as well as collision species.

\begin{figure}[thb]
\includegraphics[width=1.0\linewidth]{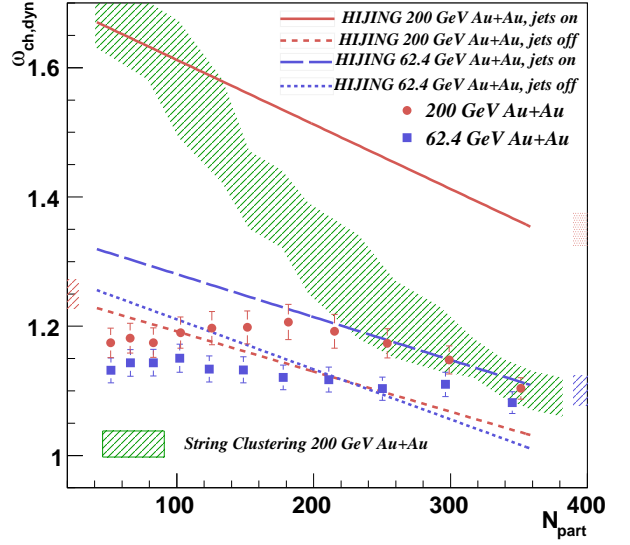} 
\includegraphics[width=1.0\linewidth]{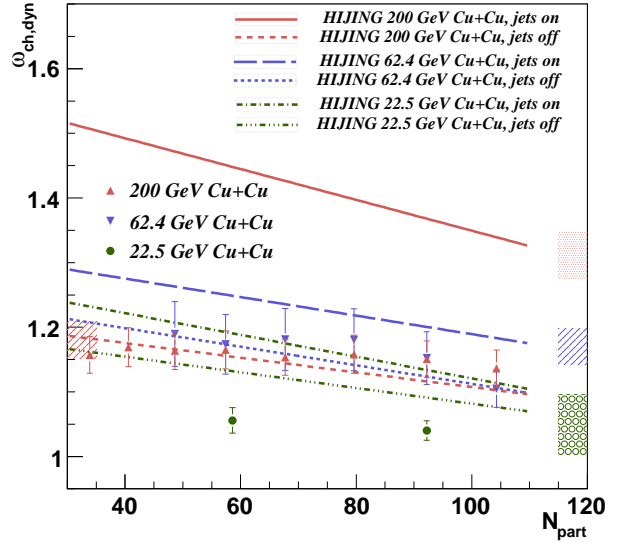} 
\caption{\label{fig:SvarHij}
Fluctuations expressed as the scaled variance as a function of 
$N_{\rm part}$ for Au+Au (upper) and Cu+Cu (lower)
collisions for $0.2<p_T<2.0$ GeV/c. The estimated 
contribution from geometry fluctuations has 
been removed. Results from the HIJING model with jets turned on and 
jets turned off are overlayed with the shaded regions representing 
the systematic error for each curve. }
\end{figure}

Baseline comparisons are facilitated by PHENIX measurements of 
charged particle multiplicity fluctuations in minimum bias 200 GeV 
p+p collisions. The p+p data and the NBD distribution to the 
multiplicity distribution are shown in Fig.~\ref{fig:nbdCuCu}. 
The NBD fit yields $\mu_{\rm ch}$ = 0.32 $\pm$ 0.003, $\omega_{\rm 
ch}$ = 1.17 $\pm$ 0.01, and $k_{\rm NBD}$ = 1.88 $\pm$ 0.01. 

\clearpage

\noindent These results are in agreement within errors with 
previous measurements in 
the same pseudorapidity range of $k_{\rm NBD} = 1.9 \pm 0.2 \pm 0.2$ 
by the UA5 Collaboration \cite{ua5200GeV} in collision of protons 
and antiprotons at 200 GeV. 
Comparisons of the participant 
superposition model to the 22.5 GeV Cu+Cu data can be made to 
multiplicity fluctuations measured in 20 GeV p+p collisions by the 
NA22 Collaboration \cite{na22Clan} over the same pseudorapidity 
range as the PHENIX Cu+Cu measurement. After scaling the NA22 scaled 
variance to the PHENIX azimuthal acceptance, the participant 
superposition model scaled variance is expected to be constant as a 
function of centrality with a value of $1.08\pm0.04$. Lacking 
multiplicity distribution data from elementary collisions at 62.4 
GeV within the PHENIX pseudorapidity acceptance, it is assumed that 
as a function of collision energy, the scaled variance in the PHENIX 
pseudorapidity acceptance scales in the same manner as in an 
acceptance of 4$\pi$, which can be parametrized from existing p+p 
and p+$\bar{p}$ data as follows \cite{ua5Clan}:
\begin{equation} \label{eq:ppMult}
   \mu_{\rm ch} \approx -4.2 + 4.69(\frac{s}{GeV^{2}})^{0.155}.
\end{equation}
Given the mean charged particle multiplicity, the scaled variance in 
p+p and p+$\bar{p}$ can be parametrized as follows \cite{heiselReview}:
\begin{equation} \label{eq:ppSvar}
   \omega_{\rm ch} \approx 0.35 \frac{(\mu_{\rm ch}-1)^{2}}{\mu_{\rm ch}}.
\end{equation}
Scaling this parametrization to match the values 
of $\omega_{\rm ch}$ at 200 GeV and 22.5 GeV, the estimated value of 
$\omega_{\rm ch}$ at 62.4 GeV is $1.15\pm0.02$.

Comparisons of the data to the predictions of the participant 
superposition model are shown in Fig.~\ref{fig:Svar} for Au+Au 
and Cu+Cu collisions. The 
shaded regions about the participant superposition model lines 
represent the systematic error of the estimates described above.  
All of the data points are consistent with or below the participant 
superposition model estimate. This suggests that the data do not 
show any indications of the presence of a critical point, where the 
fluctuations are expected to be much larger than the participant 
superposition model expectation.

\subsection{Comparisons to the HIJING model}

Shown in Fig.~\ref{fig:SvarHij} are the scaled 
variance curves from HIJING simulations into the PHENIX acceptance. 
The HIJING simulations are performed with a fixed impact parameter 
corresponding to the mean of the impact parameter distribution for 
each bin as determined by the Glauber model in order to minimize the 
geometry fluctuation component of the result. The mean and variance 
of the resulting multiplicity distributions from HIJING are 
extracted from fits to Negative Binomial Distributions. The HIJING 
simulation multiplicity fluctuations with the jet production 
parameter turned on are consistently above the data and increase 
continuously through the most peripheral collisions. This behavior 
is not consistent with the data, where the fluctuations do not 
increase in the most peripheral collisions. Although HIJING 
reproduces the total charged particle multiplicity well, it 
consistently overpredicts the amount of fluctuations in 
multiplicity. When the jet production parameter in HIJING is turned 
off, the scaled variance as a function of centrality is independent 
of collision energy, illustrating that jet production accounts for 
the energy dependence of the HIJING results. Note that the HIJING 
results with jet production turned off are in better agreement with 
the data for all collision energies. Together with the observation 
that the multiplicity fluctuations demonstrate no significant 
$p_T$-dependence, this may be an indication that correlated emission 
of particles from jet production do not significantly contribute to 
the multiplicity fluctuations observed in the data.

\subsection{Comparisons to the clan model}

The clan model \cite{clanOriginal} has been developed to interpret 
the fact that Negative Binomial Distributions describe charged 
hadron multiplicity distributions in elementary and heavy ion 
collisions. In this model, hadron production is modeled as 
independent emission of a number of hadronic clusters, $N_c$, each 
with a mean number of hadrons, $n_c$. The independent emission is 
described by a Poisson distribution with an average cluster, or 
clan, multiplicity of $\bar{N_c}$. After the clusters are emitted, 
they fragment into the final state hadrons.  The measured value of 
the mean multiplicity, $\mu_{\rm ch}$, is related to the cluster 
multiplicities by $\mu_{\rm ch} = \bar{N_c}\bar{n_c}$. In this 
model, the cluster multiplicity parameters can be simply related to 
the NBD parameters of the measured multiplicity distribution as 
follows:

\begin{equation} \label{eq:clan1}
   \bar{N_c} = k_{\rm NBD}~log(1 + \mu_{\rm ch}/k_{\rm NBD})
\end{equation}

and

\begin{equation} \label{eq:clan2}
   \bar{n_c} = (\mu_{\rm ch}/k_{\rm NBD})/log(1+\mu_{\rm ch}/k_{\rm NBD}).
\end{equation}

\begin{figure}[bht]
\includegraphics[width=1.0\linewidth]{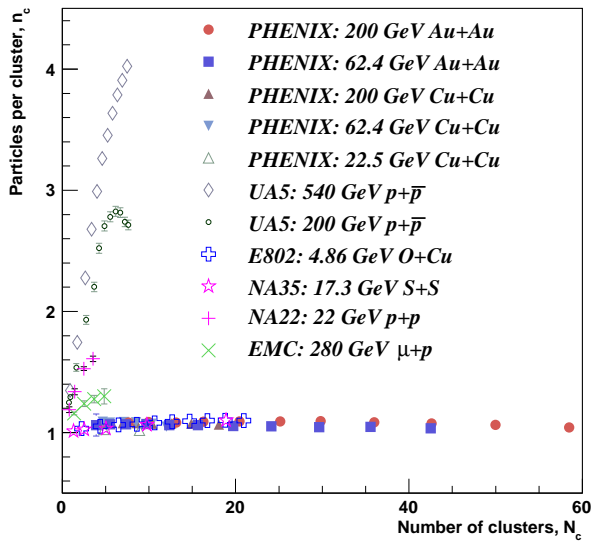} 
\caption{\label{fig:clanClusAll}
The correlation of the clan model parameters $\bar{n_c}$ and 
$\bar{N_c}$ for all of the collision species measured as a 
function of centrality. Also shown are results from 
pseudorapidity-dependent studies from elementary collisions 
(UA5~\protect\cite{ua5Clan}, EMC \protect\cite{emcClan}, and 
NA22~\protect\cite{na22Clan}) 
and heavy ion collisions 
(E802~\protect\cite{e802MF} and NA35~\protect\cite{na35Clan}). }
\end{figure}

The results from the NBD fits to the data are plotted in 
Fig.~\ref{fig:clanClusAll} for all collision species. Also shown are 
data from elementary and heavy ion collisions at various collision 
energies. The individual data points from all but the PHENIX data 
are taken from multiplicity distributions measured over varying 
ranges of pseudorapidity, while the PHENIX data are taken as a 
function of centrality. The characteristics of all of the heavy ion 
data sets are the same. The value of $\bar{n_c}$ varies little 
within the range 1.0-1.1.  The heavy ion data universally exhibit 
only weak clustering characteristics as interpreted by the clan 
model. There is also no significant variation seen with collision 
energy.  However, $\bar{n_c}$ is consistently significantly higher 
in elementary collisions. In elementary collisions, it is less 
probable to produce events with a high multiplicity, which can 
reveal rare sources of clusters such as jet production or multiple 
parton interactions.

A feature that is especially apparent in the Au+Au data is the fact 
that the scaled variance decreases with increasing centrality, with 
the most central point lying below the participant superposition 
model expectatation. The clan model provides one possible 
explanation for this effect whereby there is a higher probability 
for contributions from cluster sources such as jet production in the 
lower multiplicity peripheral events. The cluster sources introduce 
correlations that can increase the value of $1/k_{\rm NBD}$ and 
hence the value of the scaled variance of the multiplicity 
distribution. Another possible explanation for this feature can be 
addressed with a string percolation model in heavy ion collisions 
\cite{stringPerco}. In general, percolation theory considers the 
formation of clusters within a random spatial distribution of 
individual objects that are allowed to overlap with each other. The 
clusters are formed by the geometrical connection of one or more of 
the individual objects. This can be applied to estimate multiplicity 
fluctuations in heavy ion collisions whereby the objects are the 
circular cross sections of strings in the transverse plane 
\cite{mfPerco} and the strings form clusters of overlapping strings 
that then each emit a number of particles related to the number of 
strings in each cluster. As the centrality increases, the number of 
individual clusters decreases along with the variance of the number 
of strings per cluster, which can result in a decrease in the 
magnitude of the resulting multiplicity fluctuations. The prediction 
of the scaled variance from the string percolation model for 200 GeV 
Au+Au collisions scaled down to the PHENIX acceptance in azimuth and 
pseudorapidity \cite{mfPerco} is shown in Fig.~\ref{fig:SvarHij}. 
Although percolation describes the trend 
observed at the four highest centralities very well, the scaled 
variance from the model continues to increase well above the data as 
centrality decreases. The implementation of the HIJING model 
contains merging of strings that are in close spatial proximity, so 
percolation can explain the trends in the scaled variance from 
HIJING.

\section{Summary}

PHENIX has completed a survey of multiplicity fluctuations of charged 
hadrons in Au+Au and Cu+Cu collisions at three collision energies. 
The motivation for the analysis is to search for signs of a phase 
transition or the presence of the predicted critical point on the QCD 
phase diagram by looking for increased multiplicity fluctuations as a 
function of system energy and system volume. After correcting for 
non-dynamical fluctuations due to fluctuations of the collision 
geometry within a centrality bin, the multiplicity fluctuations in 
200 GeV and 62.4 GeV Au+Au collisions are consistent with or below 
the expectation from the superposition model of participant nucleons. 
The multiplicity fluctuations decrease as the collision centrality 
increases, dropping below the participant superposition model 
expectation for the most central Au+Au collisions. Fluctuations from 
Cu+Cu collisions exhibit a weaker centrality-dependence that also is 
consistent with or below the expectation from the participant 
superposition model. The absence of large dynamical fluctuations in 
excess of the participant superposition model expectation indicate 
that there is no evidence of critical behavior related to the 
compressibility observable in this dataset.  There is also no 
significant evidence of dynamical fluctuations that are dependent on 
the transverse momentum or the charge of the particles measured. As 
interpreted by the clan model, the observed fluctuations demonstrate 
only weak clustering characteristics for all of the heavy ion 
collision systems discussed here. The decreasing scaled variance with 
increasing centrality may be explained by percolation phenomena, 
however this fails to explain the most peripheral Au+Au data. 
Although this analysis does not observe evidence of critical 
behavior, it does not rule out the existence of a QCD critical point. 
Further measurements will be possible during the upcoming low energy 
scan program at RHIC allowing for a more comprehensive search for 
critical behavior.



\begin{acknowledgments}


We thank the staff of the Collider-Accelerator and Physics
Departments at Brookhaven National Laboratory and the staff of
the other PHENIX participating institutions for their vital
contributions.  We acknowledge support from the 
Office of Nuclear Physics in the
Office of Science of the Department of Energy, the
National Science Foundation, Abilene Christian University
Research Council, Research Foundation of SUNY, and Dean of the
College of Arts and Sciences, Vanderbilt University (U.S.A),
Ministry of Education, Culture, Sports, Science, and Technology
and the Japan Society for the Promotion of Science (Japan),
Conselho Nacional de Desenvolvimento Cient\'{\i}fico e
Tecnol{\'o}gico and Funda\c c{\~a}o de Amparo {\`a} Pesquisa do
Estado de S{\~a}o Paulo (Brazil),
Natural Science Foundation of China (People's Republic of China),
Ministry of Education, Youth and Sports (Czech Republic),
Centre National de la Recherche Scientifique, Commissariat
{\`a} l'{\'E}nergie Atomique, and Institut National de Physique
Nucl{\'e}aire et de Physique des Particules (France),
Ministry of Industry, Science and Tekhnologies,
Bundesministerium f\"ur Bildung und Forschung, Deutscher
Akademischer Austausch Dienst, and Alexander von Humboldt 
Stiftung (Germany),
Hungarian National Science Fund, OTKA (Hungary), 
Department of Atomic Energy (India), 
Israel Science Foundation (Israel), 
Korea Research Foundation and Korea Science and Engineering 
Foundation (Korea),
Ministry of Education and Science, Rassia Academy of Sciences,
Federal Agency of Atomic Energy (Russia),
VR and the Wallenberg Foundation (Sweden), 
the U.S. Civilian Research and Development Foundation for the
Independent States of the Former Soviet Union, the US-Hungarian
NSF-OTKA-MTA, and the US-Israel Binational Science Foundation.

\end{acknowledgments}



\begin{references}

\bibitem{QCDstephanov}
  M.~A.~Stephanov, K.~Rajagopal and E.~V.~Shuryak,
  Phys.\ Rev.\ Lett.\  {\bf 81}, 4816 (1998)
  [arXiv:hep-ph/9806219].

\bibitem{QCDstephPre}
  M.~A.~Stephanov,
  PoS {\bf LAT2006}, 024 (2006)
  [arXiv:hep-lat/0701002].

\bibitem{whitePaper}
  K.~Adcox {\it et al.}  [PHENIX Collaboration],
  Nucl.\ Phys.\  A {\bf 757}, 184 (2005)
  [arXiv:nucl-ex/0410003].

\bibitem{Stan98}
  S.~Mrowczynski,
  Phys.\ Lett.\  B {\bf 430}, 9 (1998)
  [arXiv:nucl-th/9712030].

\bibitem{Begun05}
  V.~V.~Begun, M.~I.~Gorenstein, A.~P.~Kostyuk and O.~S.~Zozulya,
  Phys.\ Rev.\  C {\bf 71}, 054904 (2005)
  [arXiv:nucl-th/0410044].

\bibitem{jeonReview}
  S.~Jeon and V.~Koch,
  arXiv:hep-ph/0304012.

\bibitem{Begun04}
  V.~V.~Begun, M.~Gazdzicki, M.~I.~Gorenstein and O.~S.~Zozulya,
  Phys.\ Rev.\  C {\bf 70}, 034901 (2004)
  [arXiv:nucl-th/0404056].

\bibitem{Becattini05}
  F.~Becattini, A.~Keranen, L.~Ferroni and T.~Gabbriellini,
  Phys.\ Rev.\  C {\bf 72}, 064904 (2005)
  [arXiv:nucl-th/0507039].

\bibitem{Stanley}
  H.~Stanley, {\it Introduction to Phase Transitions and Critical 
Phenomena} (Oxford, New York and Oxford) 1971.

\bibitem{QCDsuscept1}
  B.~J.~Schaefer and J.~Wambach,
  Phys.\ Rev.\  D {\bf 75}, 085015 (2007)
  [arXiv:hep-ph/0603256].

\bibitem{QCDsuscept2}
  C.~Sasaki, B.~Friman and K.~Redlich,
  Phys.\ Rev.\  D {\bf 75}, 054026 (2007)
  [arXiv:hep-ph/0611143].

\bibitem{droplets}
  I.~N.~Mishustin,
  Eur.\ Phys.\ J.\  A {\bf 30}, 311 (2006)
  [arXiv:hep-ph/0609196].

\bibitem{stephRajShur}
  M.~A.~Stephanov, K.~Rajagopal and E.~V.~Shuryak,
  Phys.\ Rev.\  D {\bf 60}, 114028 (1999)
  [arXiv:hep-ph/9903292].

\bibitem{kafka75}
  T.~Kafka {\it et al.},
  Phys.\ Rev.\ Lett.\  {\bf 34}, 687 (1975).

\bibitem{thome77}
  W.~Thome {\it et al.}  [Aachen-CERN-Heidelberg-Munich Collaboration],
  Nucl.\ Phys.\  B {\bf 129}, 365 (1977).

\bibitem{ua5-85}
  G.~J.~Alner {\it et al.}  [UA5 Collaboration],
  Phys.\ Lett.\  B {\bf 160}, 193 (1985).

\bibitem{ua5Clan}
  G.~J.~Alner {\it et al.}  [UA5 Collaboration],
  Phys.\ Rept.\  {\bf 154}, 247 (1987).

\bibitem{emcClan}
  M.~Arneodo {\it et al.}  [European Muon Collaboration],
  Z.\ Phys.\  C {\bf 35}, 335 (1987)
  [Erratum-ibid.\  C {\bf 36}, 512 (1987)].

\bibitem{na22Clan}
  M.~Adamus {\it et al.}  [EHS/NA22 Collaboration],
  Z.\ Phys.\  C {\bf 37}, 215 (1988).

\bibitem{ua5200GeV}
  R.~E.~Ansorge {\it et al.}  [UA5 Collaboration],
  Z.\ Phys.\  C {\bf 43}, 357 (1989).

\bibitem{e802MF}
  T.~Abbott {\it et al.}  [E-802 Collaboration],
  Phys.\ Rev.\  C {\bf 52}, 2663 (1995).

\bibitem{na35Clan}
  J.~Bachler {\it et al.}  [NA35 Collaboration],
  Z.\ Phys.\  C {\bf 57}, 541 (1993).

\bibitem{wa80MF}
  R.~Albrecht {\it et al.}  [WA80 Collaboration],
  Z.\ Phys.\  C {\bf 45}, 31 (1989).

\bibitem{wa98MF}
  M.~M.~Aggarwal {\it et al.}  [WA98 Collaboration],
  Phys.\ Rev.\  C {\bf 65}, 054912 (2002)
  [arXiv:nucl-ex/0108029].

\bibitem{na49MF}
  C.~Alt {\it et al.}  [NA49 Collaboration],
  Phys.\ Rev.\  C {\bf 75}, 064904 (2007)
  [arXiv:nucl-ex/0612010].

\bibitem{ppg061}
  S.~S.~Adler {\it et al.}  [PHENIX Collaboration],
  Phys.\ Rev.\  C {\bf 76}, 034903 (2007)
  [arXiv:0704.2894 [nucl-ex]].

\bibitem{nimPHENIX}
  K.~Adcox {\it et al.}  [PHENIX Collaboration],
  Nucl.\ Instrum.\ Meth.\  A {\bf 499}, 469 (2003).

\bibitem{nimTracking}
  K.~Adcox {\it et al.}  [PHENIX Collaboration],
  Nucl.\ Instrum.\ Meth.\  A {\bf 499}, 489 (2003).

\bibitem{nimPID}
  M.~Aizawa {\it et al.}  [PHENIX Collaboration],
  Nucl.\ Instrum.\ Meth.\  A {\bf 499}, 508 (2003).

\bibitem{nimEMC}
  L.~Aphecetche {\it et al.}  [PHENIX Collaboration],
  Nucl.\ Instrum.\ Meth.\  A {\bf 499}, 521 (2003).

\bibitem{nimReco}
  J.~T.~Mitchell {\it et al.}  [PHENIX Collaboration],
  Nucl.\ Instrum.\ Meth.\  A {\bf 482}, 491 (2002)
  [arXiv:nucl-ex/0201013].

\bibitem{nimInner}
  M.~Allen {\it et al.}  [PHENIX Collaboration],
  Nucl.\ Instrum.\ Meth.\  A {\bf 499}, 549 (2003).

\bibitem{phxMult}
  S.~S.~Adler {\it et al.}  [PHENIX Collaboration],
  Phys.\ Rev.\  C {\bf 71}, 034908 (2005)
  [Erratum-ibid.\  C {\bf 71}, 049901 (2005)]
  [arXiv:nucl-ex/0409015].

\bibitem{glauber}
  M.~L.~Miller, K.~Reygers, S.~J.~Sanders and P.~Steinberg,
  Ann.\ Rev.\ Nucl.\ Part.\ Sci.\  {\bf 57}, 205 (2007)
  [arXiv:nucl-ex/0701025].

\bibitem{heiselReview}
  H.~Heiselberg,
  Phys.\ Rept.\  {\bf 351}, 161 (2001)
  [arXiv:nucl-th/0003046].

\bibitem{phxSpec}
  S.~S.~Adler {\it et al.}  [PHENIX Collaboration],
  Phys.\ Rev.\  C {\bf 69}, 034909 (2004)
  [arXiv:nucl-ex/0307022].

\bibitem{Kon06}
  V.~P.~Konchakovski, M.~I.~Gorenstein, E.~L.~Bratkovskaya and H.~Stocker,
  Phys.\ Rev.\  C {\bf 74}, 064911 (2006)
  [arXiv:nucl-th/0606047].

\bibitem{HIJING}
  X.~N.~Wang and M.~Gyulassy,
  Phys.\ Rev.\  D {\bf 44}, 3501 (1991).

\bibitem{mfURQMD}
  B.~Lungwitz and M.~Bleicher,
  Phys.\ Rev.\  C {\bf 76}, 044904 (2007)
  [arXiv:0707.1788 [nucl-th]].

\bibitem{mfHSD}
  V.~P.~Konchakovski, M.~I.~Gorenstein and E.~L.~Bratkovskaya,
  Phys.\ Rev.\  C {\bf 76}, 031901 (2007)
  [arXiv:0704.1831 [nucl-th]].

\bibitem{mfHSDURQMD}
  V.~P.~Konchakovski, S.~Haussler, M.~I.~Gorenstein, E.~L.~Bratkovskaya, M.~Bleicher and H.~Stocker,
  Phys.\ Rev.\  C {\bf 73}, 034902 (2006)
  [arXiv:nucl-th/0511083].

\bibitem{wnm}
  A.~Bialas, M.~Bleszynski and W.~Czyz,
  Nucl.\ Phys.\  B {\bf 111}, 461 (1976).

\bibitem{stringPerco}
  M.~Nardi and H.~Satz,
  Phys.\ Lett.\  B {\bf 442}, 14 (1998)
  [arXiv:hep-ph/9805247].

\bibitem{mfPerco}
  L.~Cunqueiro, E.~G.~Ferreiro, F.~del Moral and C.~Pajares,
  Phys.\ Rev.\  C {\bf 72}, 024907 (2005)
  [arXiv:hep-ph/0505197].

\bibitem{phenixFlow}
  A.~Adare {\it et al.}  [PHENIX Collaboration],
  Phys.\ Rev.\ Lett.\  {\bf 98}, 162301 (2007)
  [arXiv:nucl-ex/0608033].

\bibitem{clanOriginal}
  A.~Giovannini and L.~Van Hove,
  Z.\ Phys.\  C {\bf 30}, 391 (1986).

\end{references}
\end{document}